\documentclass[sigconf]{acmart}

\AtBeginDocument{%
  \providecommand\BibTeX{{%
    \normalfont B\kern-0.5em{\scshape i\kern-0.25em b}\kern-0.8em\TeX}}}

\setcopyright{acmcopyright}
\copyrightyear{2018}
\acmYear{2018}
\acmDOI{10.1145/1122445.1122456}

\acmConference[Woodstock '18]{Woodstock '18: ACM Symposium on Neural
  Gaze Detection}{June 03--05, 2018}{Woodstock, NY}
\acmBooktitle{Woodstock '18: ACM Symposium on Neural Gaze Detection,
  June 03--05, 2018, Woodstock, NY}
\acmPrice{15.00}
\acmISBN{978-1-4503-XXXX-X/18/06}

\usepackage{CJK}
\usepackage{indentfirst}
\usepackage{amsmath,bm}
\usepackage{multirow}
\usepackage{booktabs}
\usepackage{graphicx}
\usepackage{subfig}
\usepackage{subfloat}
\usepackage{diagbox}

\begin{document}

\title{Recommending Accurate and Diverse Items Using Bilateral Branch Network}


\author{Yile Liang}
\affiliation{%
  \institution{Wuhan University}
  \streetaddress{16 Luojiashan Rd.}
  \city{Wuhan}
  \state{Hubei}
  \country{China}}
  \email{liangyile@whu.edu.cn}

\author{Tieyun Qian}
\affiliation{%
  \institution{Wuhan University}
  \streetaddress{16 Luojiashan Rd.}
  \city{Wuhan}
  \state{Hubei}
  \country{China}}
  \email{qty@whu.edu.cn}



\begin{abstract}
Recommender systems have played a vital role in online platforms due to the ability of incorporating users' personal tastes. Beyond accuracy, diversity has been recognized as a key factor in recommendation to broaden user's horizons as well as to promote enterprises' sales. However, the trading-off between  accuracy and diversity remains to be a big challenge, and the data and user biases have not been explored yet.

In this paper, we develop an adaptive learning framework for accurate and diversified recommendation.  We generalize recent proposed bi-lateral branch network in the computer vision community from image classification to item recommendation. Specifically, we  encode  \emph{domain level diversity} by adaptively balancing accurate recommendation in the conventional branch and diversified recommendation in the adaptive branch of a bilateral branch network. We also capture  \emph{user level diversity} using a two-way  adaptive metric learning backbone network in each branch.
We conduct extensive experiments  on three real-world datasets. Results demonstrate that our proposed approach consistently outperforms the state-of-the-art baselines.

\end{abstract}


\maketitle

\section{Introduction}
Recommender systems have been widely deployed in many web services for addressing the information overload issues.
The traditional recommendation approaches can be broadly classified into two main lines, i.e., collaborative filtering (CF) based and content based methods. CF based methods utilize historical interactions (e.g., clicks and purchases) to infer the user's personal preference and recommend items according to the matching score between the user and the target item.  Content based methods recommend items by content similarity (e.g., item attributes and text contents) between the user's historical items and the target item.

While being one of the most successful applications in online systems,  CF based recommendation favors popular items and may recommend items that are already known to the user ~\cite{AshkanKBW_ijcai15}, and content based recommendation may produce items matching the user's interests but covering a very narrow scope of topics ~\cite{QinZ13_ijcai13}. To tackle these problems, diversity-promoting (a.k.a., diversified) recommendation  has been an active research topic in recent years. It can be treated as a two-objective optimization problem, i.e., to maximize the overall relevance of a recommendation list and to minimize the similarity among the items in the list.

Diversified recommendation helps to  broaden users' horizons and increase e-commerce firms' sales \cite{ChengWMSX_www17}.
Early work in this field often  \cite{MMR_sigir98, ZhangH_recsys08, QinZ13_ijcai13, AshkanKBW_ijcai15} adopts a post-processing strategy that first generates a candidate set based on the accuracy metric and then selects a few items by maximizing the  diversity metric. Some recent approaches \cite{DPP_nips18, PD-GAN_ijcai19, 0004NLC_sigir20, DC2B_aaai20} apply determinantal point process (DPP) to modeling set diversity, and a few methods \cite{ChengWMSX_www17, BGCF_kdd20} improve recommendation diversity in the manner of end-to-end supervised learning.
Despite their effectiveness, existing diversified recommendation approaches have the following inherent limitations.

\begin{figure*}[htb]
	\centering
	\subfloat[Music\label{sfig:data_statistic_music}]{%
		\includegraphics[width=0.26\textwidth]{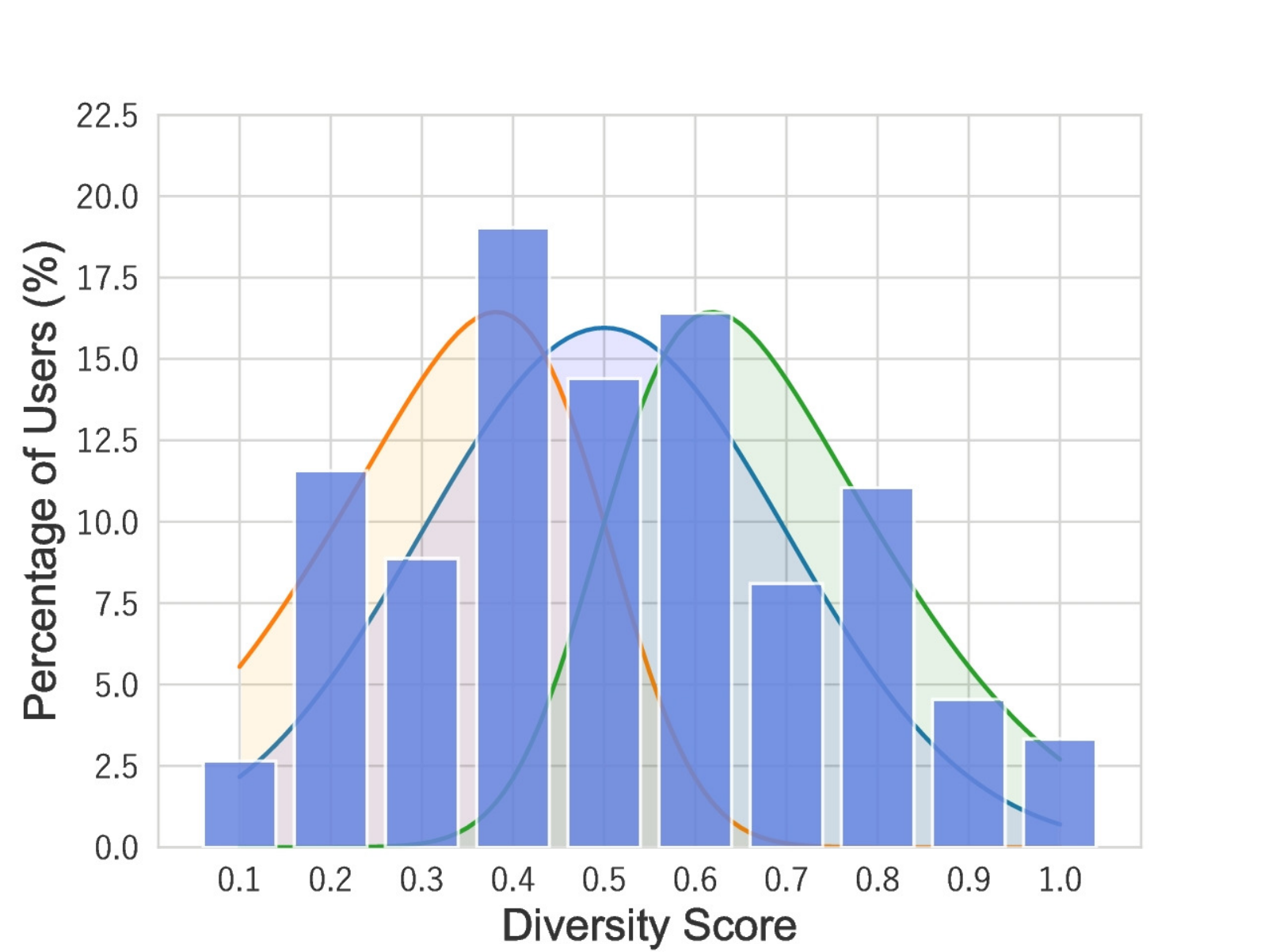}%
	}
	\subfloat[Beauty\label{sfig:data_statistic_beauty}]{%
		\includegraphics[width=0.26\textwidth]{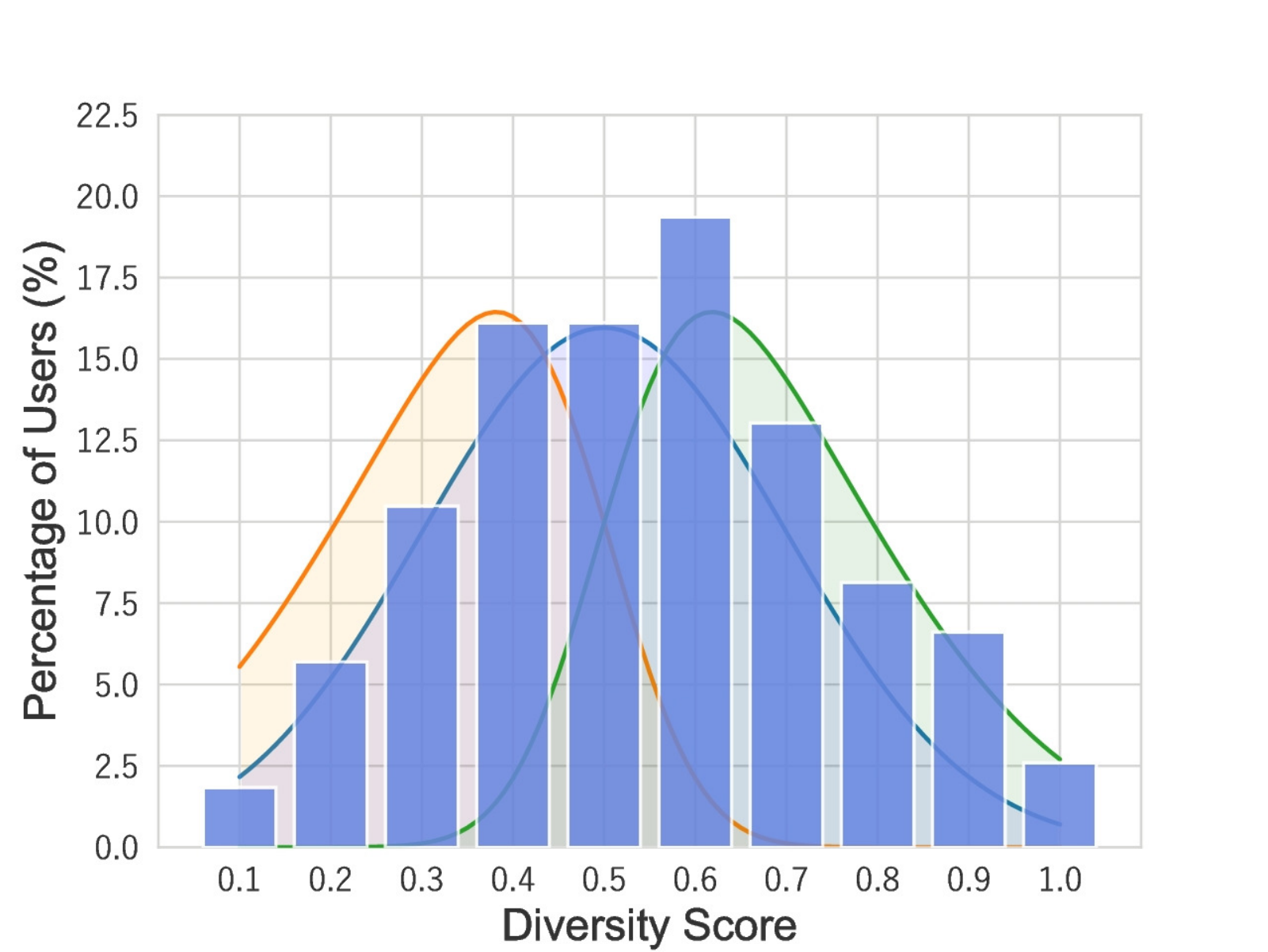}%
	}
	\subfloat[MovieLens\label{sfig:data_statistic_movielens}]{%
		\includegraphics[width=0.26\textwidth]{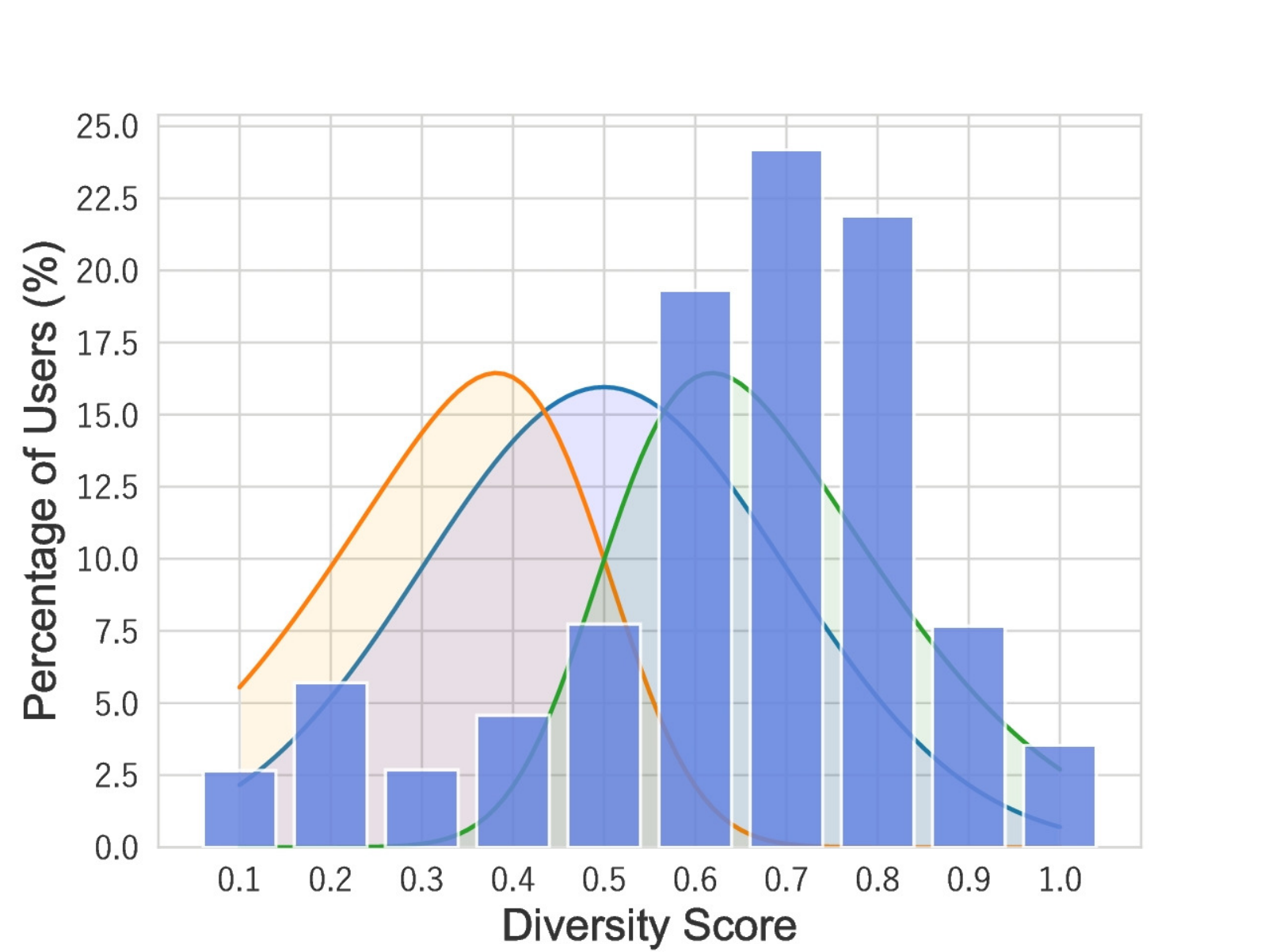}%
	}
	\vspace{-2mm}
	\caption{The distribution of users' diversity on three datasets}
	\label{fig0}
\end{figure*}

Firstly, both the post-processing and DPP based approaches \cite{QinZ13_ijcai13, AshkanKBW_ijcai15, DPP_nips18, PD-GAN_ijcai19, DC2B_aaai20} require extra parameter tuning to balance accuracy and diversity. The end-to-end method DCF \cite{ChengWMSX_www17} also needs to select a proper trade-off parameter when generating the ground-truth for training.

Secondly, none of existing work has taken care of the data and user biases. A diversity-promoting system may recommend a wide range of items to all users even if the target user has narrow and focused interests. Such a system will not change the recommendation strategy for different domains either.

Thirdly,  current studies only treat the recommendation as a task of finding items for the user, but the other side of the coin is that the products often have their own market orientations when they are designed and produced, which should not be disregarded.

\textbf{Data-Driven Study.} As an example, we plot the user's diversity distribution on three datasets in Figure \ref{fig0}. The diversity score for each user is calculated by the number of interacted categories divided by number of interacted items. The higher score, the richer diversity.
We can observe that the  distribution is skewed or central on different domains. It is interesting that a majority of users have a great deal of variability when choosing the movies' category while many users have relatively fixed preferences for the music genre. We call this \emph{domain level diversity}. We also find that each user has her own diversity preference even in the same domain. For instance, on MovieLens, a few users have narrow diversity though most of them have wide one. We call this \emph{user level diversity}.


The domain level and user level diversity reflect the characteristics of the data and the individual user, respectively, and should be presented in the recommender systems.  Equally treating all domains and users will significantly decrease recommendation performance and affect user experience. Unfortunately, none of existing methods has taken these two types of diversity into account. As a result, they are unable to tackle the problem of data and user biases.


To overcome the above limitations, we propose an adaptive learning framework with \emph{bi-lateral branch network  as the architecture} and \emph{two-way metric learning as the backbone} in each branch. Our work is inspired by the bi-lateral branch network (BBN) ~\cite{BBN_cvpr20} in visual recognition which learns representation and classifier in two branches with shared parameters, while in our model one branch is for accurate recommendation and the other for diversified recommendation and each branch has independent parameter space.

The architecture of two separate branches naturally equips the model with the ability of balancing accuracy and diversity  without extensive tuning of the trade-off parameter. We move one step further in that we propose \emph{an adaptive balancing strategy} which encodes the domain level diversity by automatically determining the order of learning focus between bilateral branches. Moreover, we design \emph{a two-way adaptive metric learning backbone network} inside each branch, which captures both the user's interests in specific items and the item's orientation towards target users, and then treats the user level diversity as a special relation connecting the user and the item.

To evaluate the recommendation performance, we conduct extensive experiments on three real-world datasets. The results demonstrate that our proposed model yields significant improvements over the state-of-the-art baselines for the trade-off performance combining both accuracy and diversity, and it also outperforms or matches the performance of these baselines in terms of single evaluation metric.

\section{Related Work}
There are two major lines of approaches in recommender systems, including collaborative filtering based recommendation and content based recommendation. Due to the space limitation, we only review the closely related literature in diversified recommendation and metric learning technique.

\subsection{Diversified Recommendation}
Diversified recommendation can be viewed from both the  aggregate and individual perspectives ~\cite{RecAdv_arxiv19}.

\textit{1) Aggregate diversity} refers to the diversity of recommendations across all users~\cite{AdomaviciusK_tkde12}. A system with high aggregate diversity will  recommend a majority number (if not all) of  items out of the entire item pool. Since the classic Pareto principle reveals that the top 20\% products often generate 80\% of sales, current work on aggregate diversity pays more attention to long-tail items. For example,  a few studies \cite{AdomaviciusK_tkde12, Oestreicher-SingerS_misq12} improve aggregate diversity by countering the effects of item popularity. Other studies \cite{Park_tkde13, YinCLYC_vldb12} propose clustering approaches and attempt to leverage long-tail items directly in a recommendation list. A recent work ~\cite{KimKPY_ijcai19} constructs a pseudo ground-truth ranking order by clustering and relocating the tailed items.

\textit{2) Individual diversity} refers to the diversity of recommendations for each target user~\cite{RecAdv_arxiv19}. The goal for individual diversity based recommendation is to achieve a balance between accuracy and diversity. Current research on individual diversity can be categorized into three groups.

The first group of methods adopts a post-processing heuristics. The classical MMR \cite{MMR_sigir98} approach represents relevance and diversity with independent metrics and maximizes the marginal relevance with a trade-off parameter. Other heuristic methods \cite{QinZ13_ijcai13, AshkanKBW_ijcai15} maximize a submodular objective function to learn the diversified model.

The second group of methods is based on the determinantal point process (DPP) which measures set diversity by describing the probability for  all subsets of the item set. DPP generates diverse lists through the maximum a posteriori (MAP) which is NP-hard and is computationally expensive even using the popular greedy algorithm. To address this issue, Chen et al. \cite{DPP_nips18} develop a novel algorithm to accelerate the MAP inference for DPP. Based on the  fast inference method ~\cite{DPP_nips18}, more recent studies \cite{PD-GAN_ijcai19, 0004NLC_sigir20, DC2B_aaai20} employ DPP to improve diversity in different recommendation tasks.

The third group of methods treats diversified recommendation as an end-to-end supervised learning task. Cheng et al. \cite{ChengWMSX_www17} propose the diversified collaborative filtering (DCF) to solve the coupled parameterized matrix factorization and structural learning problems. Sun et al. \cite{BGCF_kdd20} apply bayesian graph convolutional neural networks (BGCF) to model the uncertainty between user-item and bring diversity into recommendation indirectly.

Compared with aggregate diversity, more research attention is paid on individual diversity, which is also the focus of our paper. Despite impressive progress, both post-processing and DPP methods require great efforts in balancing the  trade-off between accuracy and diversity. Between two end-to-end methods, DCF~\cite{ChengWMSX_www17} also needs to choose a trade-off parameter, and BGCF ~\cite{BGCF_kdd20} increases diversity by adding the randomness into the data rather than explicitly modeling diversity.



\subsection{Metric Learning for Recommendation}
The goal of metric learning is to learn a distance metric that assigns smaller distances between similar pairs, and larger distances between dissimilar pairs. Metric learning has also been adopted in recommender systems ~\cite{Khoshneshin_RecSys10,LME_KDD12,PRME_ijcai15} to address the issue that inner product similarity violates the triangle inequality.

The pioneering work CML \cite{CML_www17} learns a joint user-item metric space for capturing fine-grained user preference. It minimizes the distance between each user-item interaction in Euclidean space, i.e., $d=\left\|u-v\right\|^2_2$, where $u$ and $v$ denotes  a user and an item, respectively.
To address the problem that the position of positive item may be close to the negative one,  SML \cite{SML_aaai20} proposes to  measure the trilateral relationship from both user-centric and item-centric perspectives. PMLAM \cite{PML_kdd20} is presented to generate adaptive margins for the training triples in the metric space.

Inspired by the relational metric learning in knowledge graph embedding, a number of  methods ~\cite{TransRec_recsys17, TransFM_recsys18, CTML_icdm18, TransNFCM, QianLNY_tois19, HierTrans_www20} are proposed to measure the distance between user and item with  $d=\left\|u+r-v\right\|^2_2$, where $r$ is the connective relation. For example, TransCF \cite{CTML_icdm18} constructs the translation vectors by incorporating the neighborhood information of users and items to model the heterogeneity of user-item relationships. LRML \cite{LRML_www18} introduces the memory-based attention module to learn the relation vector.  



Our framework adopts metric learning as the backbone network in each branch of BBN. Different from aforementioned translation based approaches that do not consider the diversity factor, our model aims to generate accurate and diverse recommendations. Furthermore, our model employs  two-way  ($u \rightarrow v$ and $v \rightarrow u$) translation while existing methods only perform $u \rightarrow v$ translation and ignore items' orientations towards users. 

\section{Problem Formulation and Preliminary}
\subsection{Problem Formulation}
\label{sec:porb}
Let $U = \{u_1, u_2, ..., u_M\}$ be a set of users and $V = \{v_1, v_2, ..., v_N\}$ be a set of items, where $M$ and $N$ denote the corresponding cardinalities.Let $\bm{R} \in \mathbb{R}^{M \times N}$ be the user-item interaction matrix, which indicates whether the user purchased or clicked on the item. We tackle the recommendation task with implicit feedback ~\cite{HuKV08_icdm08}, and the interaction matrix is defined as:
$$ R_{uv}=\left\{
\begin{array}{l}
1 \text{, if an interaction (user $u$, item $v$) is observed,} \\
0 \text{, otherwise.}
\end{array}
\right.
$$
The observed entries reflect users' potential interest on the item, while the unobserved entries are mixed with unknown data and negative views of the item. Given above interaction information, our main goal is to estimate users' interest for unobserved entries and rank the candidate items according to the predicted scores such that we can recommend top-$K$ preferable items for the traditional recommendation task.

Beyond that,  we particularly focus on the individual diversity in recommendation, which aims to generate diversified  recommendations for each target user. To be specific, on one hand, the recommendation quality is evaluated by a matching score between recommendation list and ground-truth list. On the other hand, the diversity is usually measured by category coverage and intra-list distance ~\cite{ZieglerMKL_www05, ParambathUG_recsys16, PD-GAN_ijcai19}.

In summary, suppose that each item belongs to one specific category, our task is to maintain the accuracy and make the recommended items away from each other and contain as many categories as possible. 

\subsection{Preliminary on Bilateral Branch Network} The basic idea of Bilateral Branch network (BBN) is originated from visual recognition tasks ~\cite{BBN_cvpr20}, where one branch retains the original characteristic and the other focuses on tail classes and two branches employ the same residual network with shared weights.  BBN takes care of  representation learning and classifier learning with a cumulative learning strategy for improving the recognition performance of long-tailed data.

In the cumulative learning component, the learning order between two branches is first  on the representation learning then to classifier learning. This is done by a controlling factor $\alpha=1-(T/T_{max})^2$ for representation branch and $1-\alpha$ for classifier branch, where $T_{max}$ and $T$ denotes the  number of total training epoches and the current epoch, respectively. $T$ increases when the training proceeds, and thus the importance of representation degrades while classifier upgrades gradually.

\section{Proposed Model}

\subsection{Model Overview}
\label{sec:model}
\begin{figure*}[htb]
	\centering
	\includegraphics[width=0.8\textwidth]{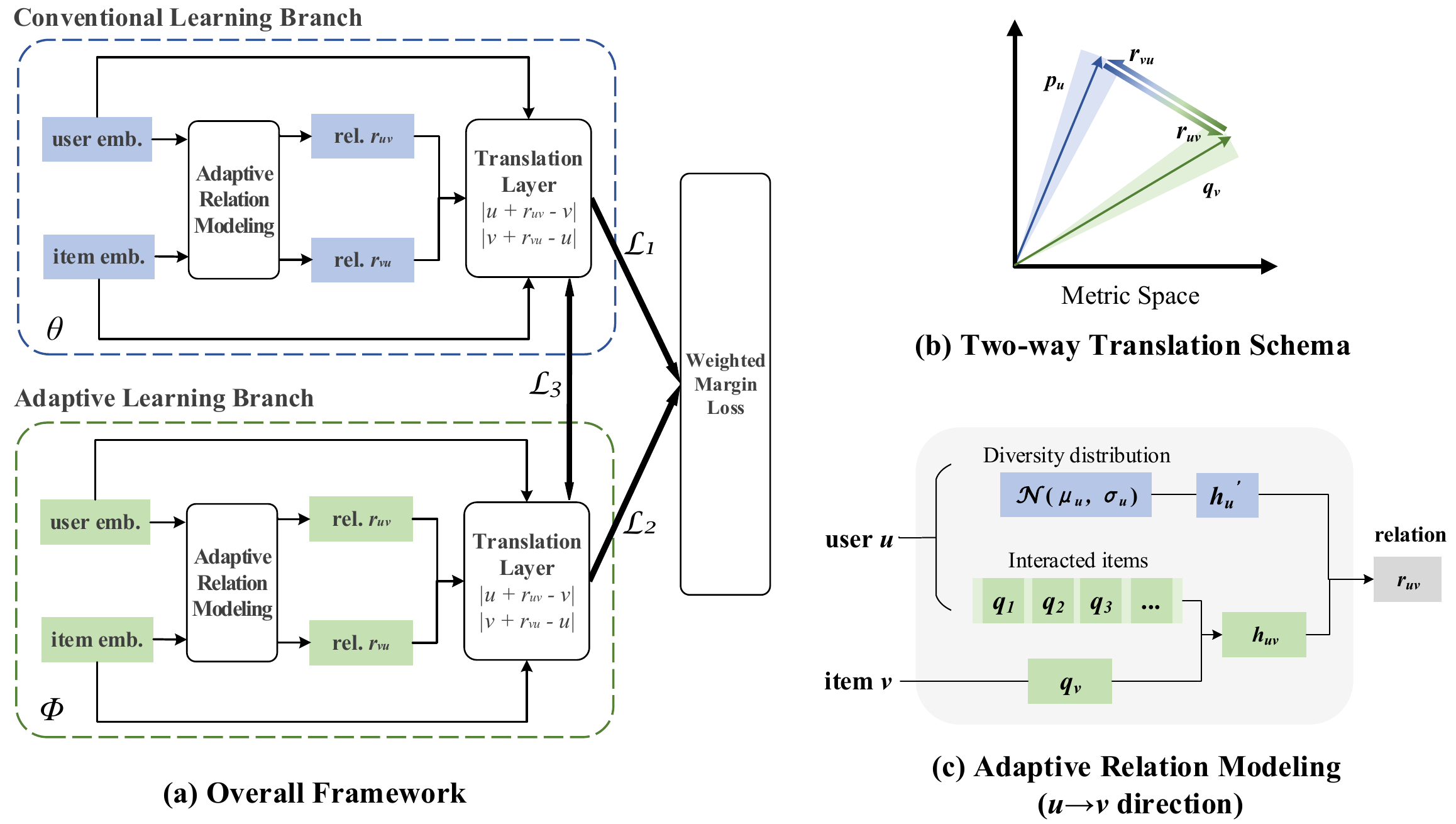} 
	\vspace{-1mm}
	\caption{An overview of our TAML model. (a) Overall framework. (b) Two-way translation scheme for modeling user-item relationship in metric space. (c) Detailed structure of adaptive relation modeling layer in the user-to-item direction. The similar process is applied to the item-to-user direction.}
	\label{fig1}
\end{figure*}

In intuition, BBN structure enables the separation of the optimization process of two objectives, where one objective is accuracy and the other is diversity in our case. In light of this, we develop a new BBN paradigm for accurate and diversified recommendation. The framework of the proposed two-way metric learning (TAML)  model is shown in Figure \ref{fig1}.

Concretely, we employ Bilateral Branch network as the main architecture, where one branch is used to improve accuracy in traditional recommendation and the other focuses on individual diversity as a subsidiary role, termed as ``conventional learning branch'' and ``adaptive learning branch'', respectively. We also present two-way adaptive metric learning with independent parameter space  as the backbone for each branch.
We here highlight two key differences between our model and the original BBN~\cite{BBN_cvpr20}.

(1) The order of learning focus in BBN is fixed, i.e., representation learning is the first and classifier learning the second. In contrast, in our task, different domains may have different distributions, and the order should be determined by the data bias rather than a fixed one. To this end, we propose an adaptive learning strategy which encodes the domain level diversity by automatically choosing the order of learning focus between bilateral branches.

(2) The backbone in each branch of BBN is the residual network which is the mainstream model in image classification, whereas in recommendation we need to measure the distance between the user and the item and each user has her own diversity preference. Hence we design a two-way adaptive metric learning backbone for the recommendation task which models the user-to-item and item-to-user relations and adaptively captures  the user level diversity simultaneously.



\subsection{Adaptively Balancing Bilateral Branches}
Existing recommender systems favor popular items and produce items matching the user's interests in a narrow scope ~\cite{QinZ13_ijcai13}. For example, if a user watched a romance film lately, the typical recommendation would be a series of redundant and similar movies. To tackle this problem, we propose to employ BBN as the main architecture, where a conventional learning branch tries to maintain the ability to recommend relevant items, and an adaptive learning branch concentrates on the items neglected by the conventional learning.
To conduct the training process on BBN, we first  choose an  order of learning focus for two branches and then feed training samples into a  corresponding branch according to different sampling strategies.

\subsubsection{Determining the order of learning focus for two branches}
\label{sec:skew}
The original BBN~\cite{BBN_cvpr20} adopts the cumulative learning strategy where the order of learning focus is from representation learning to classifier learning branch by multiplying these two branches  by a factor $\alpha$ and $1-\alpha$. This is natural for long-tailed image classification tasks as discriminative representations are the foundation for learning robust classifiers.

However, such a fixed order of focus is inappropriate for our task since users in different domains may have various diversity distributions. Essentially, the important branch should be trained later such that the network in this branch can be fine-tuned to get a better representation. For example, on MovieLens, most users have a wide range of diversity, and thus it is better to first train the conventional learning branch. If the adaptive learning branch for diversified recommendation  is trained first on this dataset, the learned embeddings will deviate from earlier diversified representations at the later periods of conventional training.
To address this issue, we propose to automatically determine the order of learning focus for balancing two branches by the domain level diversity.

We first calculate the skewness score of the distribution $X$ of users' diversity, which is defined as:
\begin{equation}\label{equ:skew}
skew(X) = E\big[(\frac{X - \mu}{\sigma})^{3}\big]
\end{equation}

A distribution with a zero skewness is a normal distribution, and a positive/negative distribution denotes a right/left skewed one. For a skewed/normal distribution, the model should place more emphasis on the adaptive/conventional learning branch. For this purpose, we associate the controlling factor $\alpha$  with the conventional learning branch while $1-\alpha$ with the adaptive learning branch for a skewed distribution, and the order will be reversed for a normal distribution. Practically, since the real world data might not strictly follow a normal distribution, we set a skew score 0.2 as a threshold for the normal distribution.

\subsubsection{Determining the sampling strategy for each branch}
\label{sec:userdiver}
In our model, the conventional learning branch is used to  keep  the accuracy of recommendation, and adaptive learning branch is for increasing the diversity. To achieve this, we apply uniform and reversed data sampler to the conventional and adaptive learning branch. Specifically, the uniform sampler samples each user-item pair with equal probability from all observed entries, and thus it can retain the characteristics of original distributions. For the reversed sampler, we incorporate domain users' diversity preference by setting the sampling possibility of each user-item pair proportional to the reciprocal of the samples in one category for the target user.

Formally, suppose that $V_{u}^{+} = \{v_1, v_2, ...\}$ is the item set interacted by a user $u$. Each item belongs to a category and $C_{u}^{+}$ is the category set of user $u$. The number of samples of category $i$ in $V_{u}^{+}$ is $N_i$ and the total number of samples in $V_{u}^{+}$ is $|V_{u}^{+}|$. We define the diversity score of the  user $u$ as $d_{u} = |C_{u}^{+}| / |V_{u}^{+}|$. The higher $d_{u}$, the more diverse interests the user has. We then construct the reversed sample set according to following steps: (1) Calculating the reversed and original sampling probability of category $i$ as:
\begin{equation}\label{equ:sampling}
P^{R}_{i} = \frac{w_{i}}{\sum_{j=1}^{|C_{u}^{+}|} w_{j}}, \quad P^{O}_{i} = \frac{N_{i}}{|V_{u}^{+}|}
\end{equation}
where $w_{i} = |V_{u}^{+}| / N_{i}$; (2) Sampling a category $i$  according to the user diversity by comparing a noise $z \sim \mathcal{U}(0,1)$ with $d_{u}$, if $z < d_u$ sampling a category using $P^{R}_{i}$ and otherwise using $P^{O}_{i}$; (3) Uniformly selecting an item belonged to category $i$ in $V_{u}^{+}$. In this way, we enforce the model to adaptively pay more or less attention to a broader or narrower range of categories of the specific user  depending on the user's diversity preference.


\subsection{Two-Way Adaptive Metric Learning inside Each Branch}
\label{sec:bml}
In each branch of BBN, we adopt a two-way adaptive metric learning backbone network. The basic idea of relational metric learning for recommendation is to measure the distance between the user and the item using the formula $d=\left\|u+r_{uv}-v\right\|^2_2$, where $r_{uv}$ is the relevance relation between $u$ and $v$. The metric learning backbone in our framework has the following distinct properties.

Firstly, unlike existing relational metric learning approaches with one-way ($u \rightarrow v$) translation, our method performs both  $u \rightarrow v$ and $v \rightarrow u$ translations.

Secondly, besides adaptively measuring the relevance relation $r_{uv}$, our method also injects the user's/item's diversity preference into translation, i.e., $r_{uv}$/$r_{vu}$ consists of a relevance relation and a diversity relation.


\subsubsection{Getting user and item embedding}
We set up a lookup table to transform the one-hot representations of each user and item into low-dimensional vector in the metric space. After transformation, $\bm{p}_u \in \mathbb{R}^{M \times D}, \bm{q}_v \in \mathbb{R}^{N \times D}$ denotes the latent factor representation of the user $u$ and the item $v$, respectively.

\subsubsection{Adaptively measuring the relevance relation}
We design a novel attentive matching method to adaptively measuring the relevance relation between user $u$ and item $v$ in the metric space.
Given the user-item pair $(u,v)$, we learn the relevance relation vector $\bm{h}_{uv}$ and $\bm{h}_{vu}$ by selecting relevant pieces of historical behaviors. Suppose that $V_{u}^{+} = \{v_1, v_2, ...,v_{L_u}\}$ are interacted items of user $u$, the the relevance relation performs as a weighted sum to adaptively aggregate representation:
\begin{equation}\label{equ:attn}
\bm{h}_{uv} = f(\bm{p}_{u}, \bm{q}_{v_1}, ..., \bm{q}_{v_{L_u}}) = \sum_{j \in V_{u}^{+}} a_{u,j} \bm{q}_{j}.
\end{equation}
Each element of the attention vector $\bm{a}_{u}$ can be defined as:
\begin{equation}\label{equ:attn2}
a_{u,j} = softmax(\bm{p}_u^{\top} \bm{W}_a \bm{q}_j),
\end{equation}
where $\bm{W}_a \in \mathbb{R}^{D \times D}$ is the weight matrix to be learned, and the $softmax(.)$ is applied to the original attention weights for  normalization. Similar operation can be conducted by extracting relevant information from interacted users of item $v$ as:
\begin{equation}\label{equ:attn3}
\bm{h}_{vu} = f(\bm{q}_{v}, \bm{p}_{u_1}, ..., \bm{p}_{u_{L_v}}) = \sum_{j \in U_{v}^{+}} a_{v,j} \bm{p}_{j},
\end{equation}
where $U_{v}^{+} = \{u_1, u_2, ...,u_{L_v}\}$ is the user set interacted by item $v$.

\subsubsection{Adaptively measuring the diversity relation}
Apart from the relevance relation between user $u$ and item $v$, we believe the user's preference to diversity would affect the translation process, e.g., users having broader interests are more likely to watch various types of movies. The same goes for the item. In view of this, we treat the user's/item's preference to diversity as an additional diversity relation $\bm{h}'_{u}$/$\bm{h}'_{v}$ to connect $u$ and $v$, and represent it as a Gaussian distribution for each user or item, i.e.,
\begin{equation}\label{equ:gaus}
\begin{split}
\bm{h}'_{u} \sim \mathcal{N}(\bm{\mu}_u, \bm{\sigma}_u), \\
\bm{h}'_{v} \sim \mathcal{N}(\bm{\mu}_v, \bm{\sigma}_v),
\end{split}
\end{equation}
where $\bm{\mu}_u(\bm{\mu}_v)$ and $\bm{\sigma}_u(\bm{\sigma}_v)$ are learned mean vector and standard deviation vector for user $u$ (item $v$), respectively. To perform back-propagation from  $\bm{h}'$, we adopt the reparameterization trick ~\cite{VAE_ICLR14}. We sample $\bm{\epsilon} \sim \mathcal{N}(0, \bm{I})$ and reparameterize $\bm{h}' = \bm{\mu} + \bm{\epsilon} \odot \bm{\sigma}$, where $\odot$ denotes element-wise product.

Note that, the parameters $\bm{\mu}, \bm{\sigma}$ are not item-specific or user-specific, because it is hard to learn the characteristics from interactions directly. Instead, we relate these parameters to coarse-grained aspects, which are more suitable for modeling the variance among clusters of items. Specifically, we first calculate the frequency vector from each user to $|C|$ categories from the interaction matrix. After dimension reduction and normalization, we obtain the user's attention distribution $\bm{w}_u$ over $K$ aspects. We then construct the mean and standard deviation vector as:
\begin{equation}\label{equ:gaus2}
\begin{split}
\bm{\mu}_{u} = \bm{T}_{\mu}^\top \bm{w}_u, \\
\bm{\sigma}_{u} = \bm{T}_{\sigma}^\top \bm{w}_u,
\end{split}
\end{equation}
where $\bm{T}_{\mu} \in \mathbb{R}^{K \times D}$ and $\bm{T}_{\sigma} \in \mathbb{R}^{K \times D}$ are the learnable weight matrix shared for all users. Similarly, $\bm{\mu}_v$ and $\bm{\sigma}_v$ can be constructed in the similar way.

\subsubsection{Performing two-way translation}
So far, we have extracted two types of relations accounting for the relevance and diversity. We now incorporate these relations into the final connective relation as:
\begin{equation} \label{equ:rel}
\begin{split}
\bm{r}_{uv} = \bm{h}_{uv} + \bm{h}'_u, \\
\bm{r}_{vu} = \bm{h}_{vu} + \bm{h}'_v.
\end{split}
\end{equation}

Then we design the distance function for each user-item interaction. As we pointed out in previous section, classic one-way translation based metric learning has limited expression ability. Indeed, not only a user chooses preferable items according to her own interests, but also an item is oriented towards suitable consumers by its characteristic. Hence we perform a two-way translation in our model. Given the user-item pair $(u, v)$, the distance function can be defined as:
\begin{equation} \label{equ:trans}
\begin{split}
d(u,v) = \left\|\bm{p}_u + \bm{r}_{uv} - \bm{q}_v\right\|^2_2, \\
d(v,u) = \left\|\bm{q}_v + \bm{r}_{vu} - \bm{p}_u\right\|^2_2,
\end{split}
\end{equation}
where $\left\|.\right\|_2$ is the L2 norm. The two-way translation improves the expression ability from two separate perspectives compared to the vanilla metric learning approaches.

\subsection{Optimization and Learning}
\label{sec:opt}
Under the BBN architecture, two types of training samples are fed into corresponding branches to calculate the loss. We define $\mathcal{L}_{B1}, \mathcal{L}_{B2}$ as the loss for conventional learning branch  and adaptive learning branch, respectively. We extend the pairwise margin loss ~\cite{CML_www17} for two-way adaptive metric learning in each branch, which ensures the distance between user $u$ and positive item $v$ is less than the distance between user $u$ and negative
item $v^-$ by a fixed margin $m > 0$. The loss is defined as follows:
\begin{equation} \label{equ:loss}
\begin{aligned}
\mathcal{L}_{Bi} = \sum_{(u,v) \in \mathcal{I}}\sum_{(u,v^-) \in \mathcal{I}^-} \left[d(u,v) - d(u,v^-) + m\right]_{+} \\
+ \left[d(v,u) - d(v^-,u) + m\right]_{+},
\end{aligned}
\end{equation}
where $(u,v) \in \mathcal{I}$ indicates the positive sample, and $(u,v^-) \in \mathcal{I}^-$ denotes the negative sample and $\left[x\right]_+ = max(x,0)$.

In addition, since the translations in two branches are conducted in different parameter spaces, there might be a small gap between them.  We then define an additional consistency loss $\mathcal{L}_3$ to measure the distribution difference between two branches:
\begin{equation} \label{equ:loss2}
\mathcal{L}_{3} = \mathcal{D}_{KL}\left(U_{fw}(x) || R_{fw}(x')\right) + \mathcal{D}_{KL}\left( U_{bw}(x) || R_{bw}(x') \right) ,
\end{equation}
where $\mathcal{D}_{KL}$ is Kullback-Leibler divergence, $U(x)$ and $R(x')$ denote the probability distributions obtained by calculating the softmax function over the translations from uniform and adaptive learning branches, respectively. The subscript ``\textit{fw}'' and ``\textit{bw}'' denote user-to-item direction and item-to-user direction, respectively.  

Finally, we define the overall loss function $\mathcal{L}$ as:
\begin{equation} \label{equ:loss3}
\mathcal{L}=
\begin{cases}
\alpha \mathcal{L}_{B1} + (1-\alpha)\mathcal{L}_{B2} + \mathcal{L}_3, & \text{if\ } \ skew < 0.2,\\
(1-\alpha) \mathcal{L}_{B1} + \alpha \mathcal{L}_{B2} + \mathcal{L}_3,  & \text{if\ } \ skew \ge 0.2.
\end{cases}
\end{equation}
where $\alpha$ is the parameter to control the weights of two branches, which is calculated by $d_{u} * T / T_{max}$ to include the individual user's diversity preference, where $d_{u}$ is the diversity score of user $u$ as defined in Section~\ref{sec:userdiver}, and $T$ and $T_{max}$ are the current and total training epoch number. Also note that either $\alpha$ or $1-\alpha$ can be associated with $\mathcal{L}_{B1}$ or $\mathcal{L}_{B2}$, depending on the domain has a skewed or normal diversity distribution.


\section{Experiments}
In this section, we design experiments to verify the effectiveness and correctness of our model. We begin with the experimental setup, and then analyze the experimental results by comparing it with the state-of-the-art baselines. We drill down to the model details with both quantitative experiments and qualitative studies.

\subsection{Experimental Setup}
\subsubsection{Datasets}
\begin{table}[htb]
	\caption{Statistics of the datasets.}
	\label{tab:dataset}
	\centering
	\renewcommand\arraystretch{1.2}
	\begin{tabular}{|c|c|c|c|c|c|}
		\hline
		Datasets & \#Users & \#Items & \#Interactions &Sparsity &\#Cate.\\ \hline \hline
		Music & 5541 & 3568 & 64786 & 99.67\% & 60 \\ \hline
		Beauty & 8159 & 5862 & 98566 & 99.79\% & 41 \\ \hline
		Movielens & 6040 & 3706 & 1000209 & 95.53\% & 18\\ \hline
	\end{tabular}
\end{table}

We use three publicly available datasets from different domains.
\textbf{Music} and \textbf{Beauty} are chosen from Amazon\footnote{http://jmcauley.ucsd.edu/data/amazon/links.html} which both contain metadata of diverse products. Following the evaluation settings in ~\cite{A3NCF_ijcai18, Chen_NARRE_www2018}, we take the 5-core version for experiments, where each user or item has at least five interactions.
\textbf{MovieLens}\footnote{https://grouplens.org/datasets/movielens/} is a widely adopted dataset in the application domain of recommending movies to users. We employ the MovieLens-1M version.

In order to be consistent with the implicit feedback setting~\cite{NCF_www17, enmf_tois20}, we transform the the detailed rating into a value of 0 or 1, indicating whether a user has rated an item. The statistics of the datasets are shown in Table \ref{tab:dataset}.

\subsubsection{Evaluation Metrics}
For all experiments, we evaluate the recommendation performance in terms of accuracy and diversity.

\textit{Accuracy.} We evaluate the accuracy of ranking list using Recall ~\cite{CML_www17} and Normalized Discounted Cumulative Gain (NDCG) ~\cite{NCF_www17,enmf_tois20}. Recall considers whether the ground-truth is ranked amongst the top K items while NDCG is a position-aware metric, which assigns higher score to the hits at higher positions.

\textit{Diverisity.} We evaluate the recommendation diversity by the
intra-list distance (ILD) ~\cite{ZhangH_recsys08, ParambathUG_recsys16} and category coverage (CC) ~\cite{ParambathUG_recsys16,PD-GAN_ijcai19}. ILD measures the diversity of the recommended item set by the mean distance between all pairs of items, while CC measures the coverage of the user interests by counting the relevant categories in the set.

\textit{Trade-off.} Following ~\cite{ChengWMSX_www17, DC2B_aaai20},  we employ F-score as a harmonic mean of conflicting accuracy and diversity for different methods, where $F$-$score=2*accuracy*diversity / (accuracy+diversity)$.

\subsubsection{Baselines}
To demonstrate the effectiveness of our proposed TAML model, we compare it with the following methods.

\begin{itemize}
	\item \textbf{LFM} ~\cite{MFT_com09} is a classical matrix factorization method for collaborative filtering, which learns the latent factors by alternating least squares.
	
	\item \textbf{NCF} ~\cite{NCF_www17} presents a neural collaborative filtering method combining multi-layer perceptron with generalized matrix factorization to encode non-linearities.
	
	\item \textbf{CML} ~\cite{CML_www17} is a metric learning method, which encodes the user and item into a joint metric space and measures the user-item pair by euclidean distance.
	
	\item \textbf{TransCF} ~\cite{CTML_icdm18}  is a relational metric learning method which includes the neighborhood information of users and items to construct	translation vectors. 
	
	\item \textbf{ENMF} ~\cite{enmf_tois20} proposes to learn neural recommendation model from the whole training data with a non-sampling strategy to enhance learning efficiency.
	
	\item \textbf{MMR} ~\cite{MMR_sigir98}  is a canonical re-ranking method for diversified ranking problems by  maximizing the marginal relevance.
	
	\item \textbf{DPP} ~\cite{DPP_nips18} applies determinantal point process to optimize the trade-off between accuracy and diversity and generates recommendation list through the MAP inference.
	
	\item \textbf{PD-GAN} ~\cite{PD-GAN_ijcai19} proposes to learn users' personal preferences and item diversity through an adversarial learning process and adopts DPP model as generator to generate diverse results.
	
	\item \textbf{BGCF} ~\cite{BGCF_kdd20} models the uncertainty in the user-item bipartite graph by node copying and achieves accurate and diverse results using Bayesian graph convolutional neural networks.
\end{itemize}

We partition the baselines into two parts. The first part (from LFM to ENMF) includes the classical methods designed for improving the accuracy. The second part (from MMR to BGCF) includes typical diversity-promoting methods (post-processing, DPP, and end-to-end) developed for balancing accuracy and diversity.

\subsubsection{Settings}
For each user, we randomly select 80\% of historical interactions of each user as training set, and the remaining 20\% data constitutes the testing set. During the test phrase, our evaluation protocol ranks all unobserved items in the training set for each user \cite{EHCF_aaai20}. 

For the baselines, we follow the same hyper-parameter settings if they are reported by the authors and we fine-tune the hyper-parameters if they are not reported. We use WMF ~\cite{HuKV08_icdm08} to calculate the relevance score in first step for MMR and to construct kernel matrix for DPP. The trade-off parameter in MMR, DPP, and PD-GAN is set to 0.8, 0.9, and 0.9, respectively for getting the best trade-off performance.  For BGCF, we set the number of sampled graphs to 5.

For our TAML,  we set the batch size = 128, the number of max epoch = 20, the initial learning rate = 0.0005, and we use Adam ~\cite{KingmaB14} as optimizer to self-adapt the learning rate. The margin $m$ in margin loss = 1. The embedding dimension $D$ = 50. The number of aspects $K$ = 20. We sample $P$ = 20 unobserved items as negative samples for each positive user-item pair ~\cite{CML_www17}. 

\begin{table*}[]
	\caption{The overall performance comparison on three datasets in terms of accuracy, diversity and trade-off evaluation, respectively. The best performance among all is in bold while the second best one is marked with an underline. }
	\label{tab:performance}
	\centering
    \small
	\setlength{\tabcolsep}{1.96mm}
	\def\arraystretch{1.05}
	\begin{tabular}{|c|c|c|c|c|c|c|c|c|c|c|c|}
		\hline
		\multirow{2}{*}{Datasets}  & \multirow{2}{*}{Models} & \multicolumn{4}{c|}{Acurracy}                                         & \multicolumn{4}{c|}{Diversity}                                        & \multicolumn{2}{c|}{Trade-off}    \\ \cline{3-12}
		&                         & Recall@5        & Recall@10       & NDCG@5          & NDCG@10         & ILD@5           & ILD@10          & CC@5            & CC@10           & F1@5            & F1@10           \\ \hline \hline
		\multirow{9}{*}{Music}     & LFM                     & 0.1264          & 0.1822          & 0.1273          & 0.1503          & 0.5904          & 0.6237          & 0.5762          & 0.6820          & 0.2082          & 0.2820          \\
		& NCF                     & 0.1156          & 0.1668          & 0.1151          & 0.1364          & 0.6556          & 0.6809          & 0.5814          & 0.6897          & 0.1965          & 0.2680          \\
		& CML                     & 0.1579    & 0.2204    & \underline{0.1645}    & \underline{0.1879}    & 0.5871          & 0.6366          & 0.5889          & 0.6892          & \underline{0.2489}    & \underline{0.3274}    \\
		& TransCF & \underline{0.1598}   & \underline{0.2242}    & 0.1611 & 0.1871  & 0.5377    & 0.5801 & 0.5920 & 0.6809 & 0.2464    & 0.3234 \\
		& ENMF                    & 0.1560          & 0.2179          & 0.1559          & 0.1796          & 0.5960          & 0.6418          & 0.5854          & 0.6917          & 0.2473          & 0.3253          \\ \cline{2-12}
		& MMR                     & 0.0690          & 0.1029          & 0.0708          & 0.0860          & \textbf{0.7557} & \textbf{0.7822} & 0.6085          & \textbf{0.7273} & 0.1265          & 0.1819          \\
		& DPP                     & 0.0771          & 0.1528          & 0.0772          & 0.1102          & \underline{0.6769}    & 0.6800          & \underline{0.5926}    & 0.7136          & 0.1384          & 0.2495          \\
		& PD-GAN                  & 0.1435          & 0.2068          & 0.1401          & 0.1658          & 0.6030          & 0.6376          & 0.5732          & 0.6806          & 0.2318          & 0.3123          \\
		& BGCF                    & 0.1340          & 0.1934          & 0.1376          & 0.1621          & 0.6023          & 0.6246          & 0.5835          & 0.6870          & 0.2192          & 0.2953          \\ \cline{2-12}
		& Ours                    & \textbf{0.1685} & \textbf{0.2327} & \textbf{0.1718} & \textbf{0.1960} & 0.6412          & \underline{0.6893}    & \textbf{0.6135} & \underline{0.7168}    & \textbf{0.2669} & \textbf{0.3479} \\ \hline \hline
		\multirow{9}{*}{Beauty}    & LFM                     & 0.0505          & 0.0781          & 0.0626          & 0.0755          & 0.7452          & 0.7510          & 0.4820          & 0.6132          & 0.0946          & 0.1415          \\
		& NCF                     & 0.0399          & 0.0654          & 0.0531          & 0.0657          & 0.7498          & 0.7719          & 0.4542          & 0.5864          & 0.0758          & 0.1206          \\
		& CML                     & 0.0605          & 0.0977          & 0.0792          & 0.0949          & 0.7336          & 0.7513          & 0.4823          & 0.6133          & 0.1118          & 0.1729          \\
		& TransCF & 0.0621   & 0.0970    & 0.0763 & 0.0927  & 0.6934    & 0.7130 & 0.4901 & 0.6125 & 0.1140    & 0.1708 \\
		& ENMF                    & \underline{0.0675}    & \underline{0.1037}    & \underline{0.0873}    & \underline{0.1042}    & 0.7084          & 0.7257          & \underline{0.4964}    & 0.6155          & \underline{0.1233}    & \underline{0.1815}    \\ \cline{2-12}
		& MMR                     & 0.0424          & 0.0791          & 0.0557          & 0.0714          & 0.7450          & 0.7509          & 0.4704          & 0.5904          & 0.0802          & 0.1431          \\
		& DPP                     & 0.0382          & 0.0800          & 0.0500          & 0.0696          & \textbf{0.7785} & \underline{0.7854}    & 0.4850          & \underline{0.6350}    & 0.0728          & 0.1452          \\
		& PD-GAN                  & 0.0580          & 0.0899          & 0.0744          & 0.0881          & 0.7309          & 0.7487          & 0.4844          & 0.6056          & 0.1075          & 0.1605          \\
		& BGCF                    & 0.0525          & 0.0867          & 0.0662          & 0.0824          & 0.7492          & 0.7524          & 0.4846          & 0.6181          & 0.0981          & 0.1555          \\ \cline{2-12}
		& Ours                    & \textbf{0.0721} & \textbf{0.1071} & \textbf{0.0926} & \textbf{0.1067} & \underline{0.7675}    & \textbf{0.7923} & \textbf{0.5009} & \textbf{0.6445} & \textbf{0.1318} & \textbf{0.1887} \\ \hline \hline
		\multirow{9}{*}{Movielens} & LFM                     & 0.0835          & 0.1391          & 0.6106          & 0.6303          & 0.7928          & 0.8048          & 0.3670          & 0.5237          & 0.1511          & 0.2372          \\
		& NCF                     & 0.0855          & 0.1431          & 0.6041          & 0.6260          & 0.7636          & 0.7798          & 0.3678          & 0.5178          & 0.1538          & 0.2418          \\
		& CML                     & \underline{0.0953}    & \underline{0.1578}    & \underline{0.6488} & \underline{0.6634} & 0.7868          & 0.8012          & \underline{0.3745}    & \underline{0.5309}    & \underline{0.1700}    & \underline{0.2637}    \\
		& TransCF & 0.0939   & 0.1562    & 0.6372 & 0.6555  & 0.7699    & 0.7869 & 0.3655 & 0.5180 & 0.1674    & 0.2607 \\
		& ENMF                    & 0.0928          & 0.1546          & 0.6131          & 0.6355          & 0.7737          & 0.7863          & 0.3700          & 0.5231          & 0.1657          & 0.2584          \\ \cline{2-12}
		& MMR                     & 0.0435          & 0.0798          & 0.4125          & 0.4571          & 0.7950          & 0.8033          & 0.3521          & 0.5112          & 0.0825          & 0.1452          \\
		& DPP                     & 0.0594          & 0.1071          & 0.4643          & 0.5096          & \underline{0.8157}    & \underline{0.8131}    & 0.3739          & 0.5126          & 0.1107          & 0.1893          \\
		& PD-GAN                  & 0.0849          & 0.1443          & 0.5705          & 0.6006          & 0.7577          & 0.7750          & 0.3616          & 0.5092          & 0.1527          & 0.2433          \\
		& BGCF                    & 0.0744          & 0.1257          & 0.5749          & 0.6004          & 0.8022          & 0.8117          & 0.3643          & 0.5243          & 0.1362          & 0.2177          \\ \cline{2-12}
		& Ours                    & \textbf{0.0975} & \textbf{0.1613} & \textbf{0.6509}    & \textbf{0.6684}    & \textbf{0.8672} & \textbf{0.8735} & \textbf{0.3901} & \textbf{0.5651} & \textbf{0.1753} & \textbf{0.2723} \\ \hline
	\end{tabular}
\end{table*}

\subsection{Performance Comparison}
The performance comparison between our proposed TAML model and the baselines on three datasets is reported in Table \ref{tab:performance}. From the results, we have the following important observations.

Firstly, it is clear that our TAML significantly outperforms all the baselines in terms of accuracy and diversity trade-off (F1) on three datasets. The relative improvement of our model over the best baselines on three datasets  are 5.58\%, 4.87\%, and 2.31\% on F1@5, and 4.34\%,  3.14\%, and 2.35\% on F1@10, respectively.
	In addition, TAML achieves the best or second best recommendation accuracy (Recall, NDCG) and diversity (ILD, CC) in most of cases. It verifies the superiority of our proposed framework with the adapted BBN architecture and two-way adaptive metric learning backbone.
	
Secondly, compared with four accuracy-oriented baselines, our TAML achieves the superior or competitive performance in terms of accuracy, and gains remarkable improvements on diversity. Among these baseline methods, LFM and NCF have inferior performance on three datasets. This indicates that the latent features obtained by matrix factorization  are insufficient to capture user and item relationships and lead to limited performance. CML and ENMF perform better, owing to that CML overcomes the inherent limitation of the inner product and ENMF uses whole-data learning strategy. However, these methods focus on the recommendation accuracy and rely on the historical interactions to learn representation, resulting in redundant recommendation results.
	
Thirdly, compared with four diversity-promoting baselines, our TAML gets significant improvements in accuracy while keeping a comparable or better diversity performance. In contrast, MMR and DPP maintain the explicit trade-off between accuracy and diversity, but they make great sacrifice of accuracy to generate a high diversity. The end-to-end methods PD-GAN and BGCF slightly enhance recommendation accuracy with a relatively low diversity. 

\subsection{Parameter Analysis}
\begin{figure*}[htb]
	\vspace{-2mm}
	\centering
	\subfloat[Music\label{sfig:music_dimension}]{%
		\includegraphics[width=0.3\textwidth]{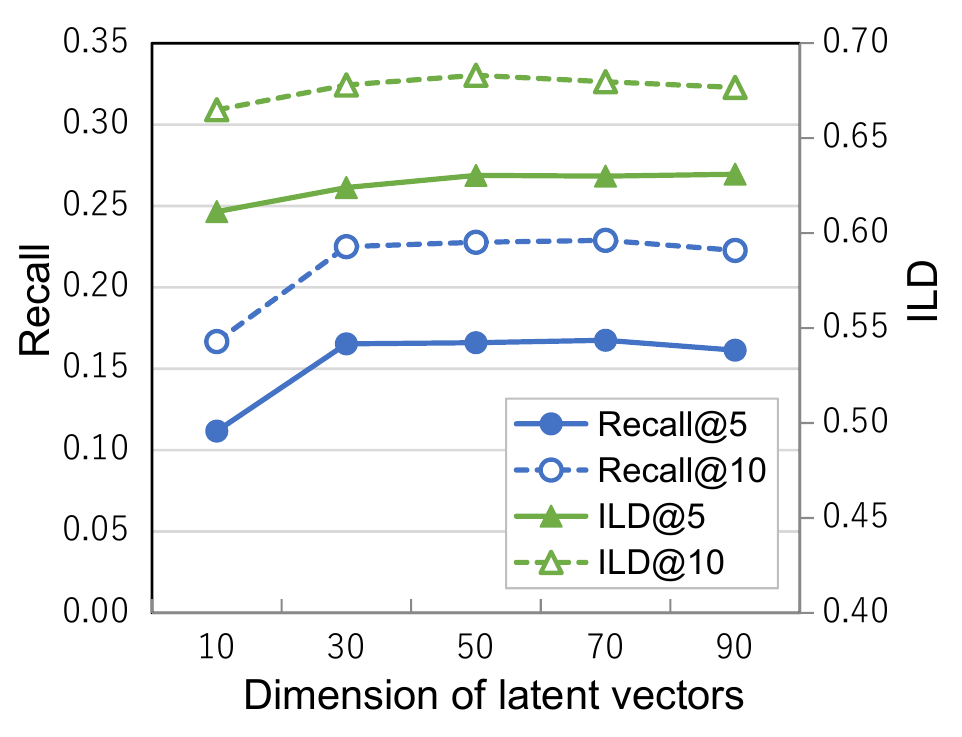}%
	}
	\subfloat[Beauty\label{sfig:beauty_dimension}]{%
		\includegraphics[width=0.3\textwidth]{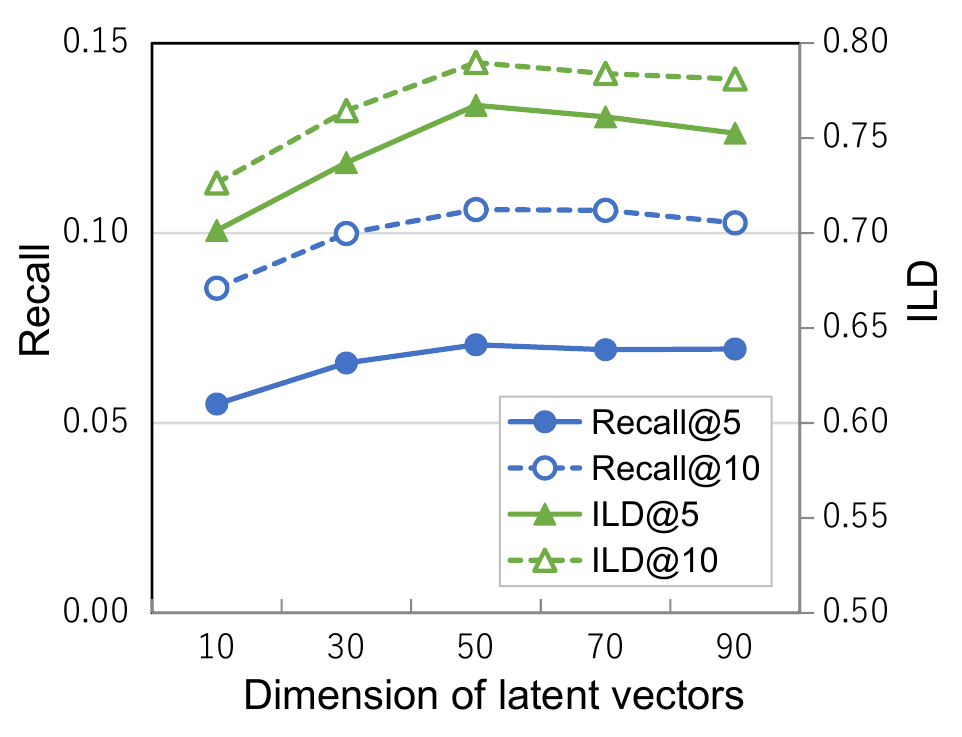}%
	}
	\subfloat[Movielens\label{sfig:movielens_dimension}]{%
		\includegraphics[width=0.3\textwidth]{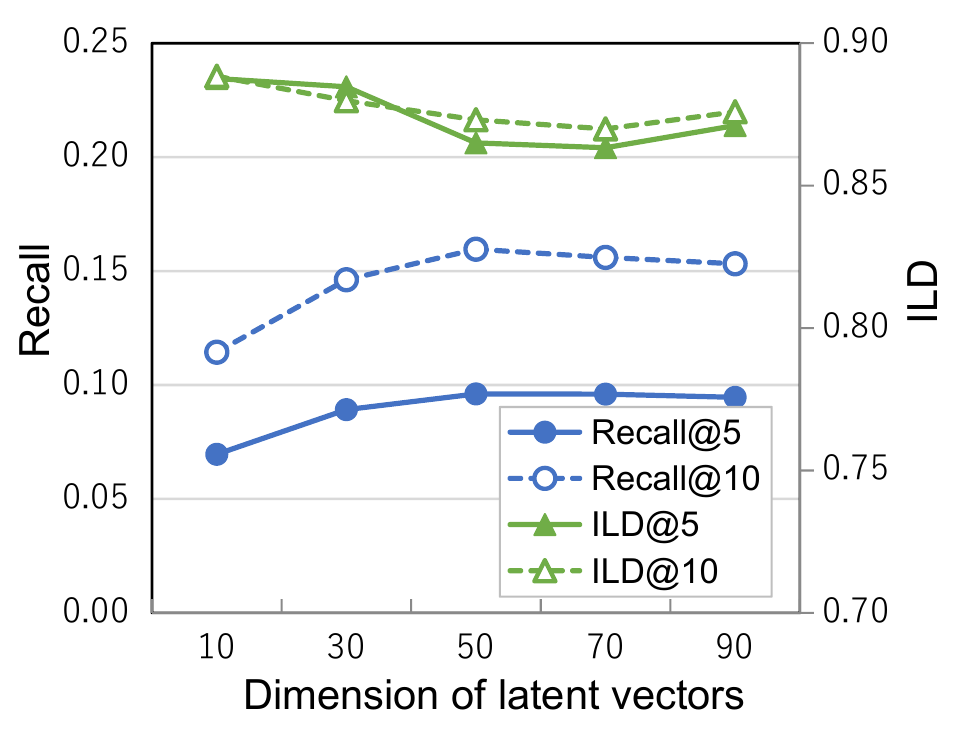}%
	}
	\vspace{-0mm}
	\caption{Impacts  of latent vector dimension $D$.}
	\label{fig:dim_variation}
	\vspace{-1mm}
\end{figure*}

\begin{figure*}[htb]
	\vspace{-2mm}
	\centering
	\subfloat[Music\label{sfig:music_negative}]{%
		\includegraphics[width=0.3\textwidth]{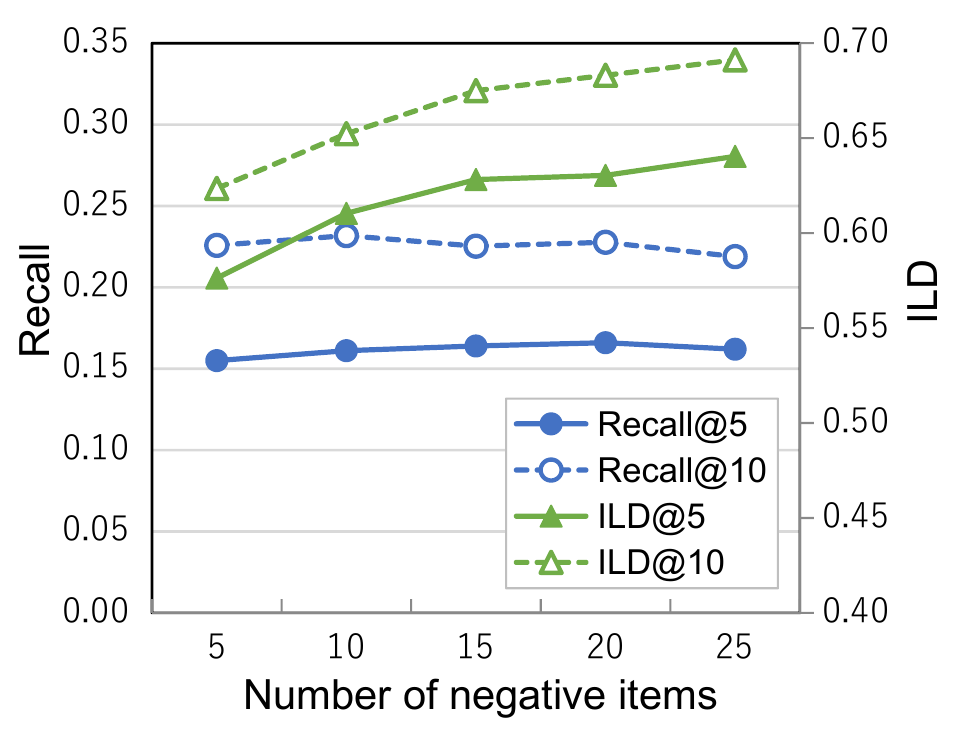}%
	}
	\subfloat[Beauty\label{sfig:beauty_negative}]{%
		\includegraphics[width=0.3\textwidth]{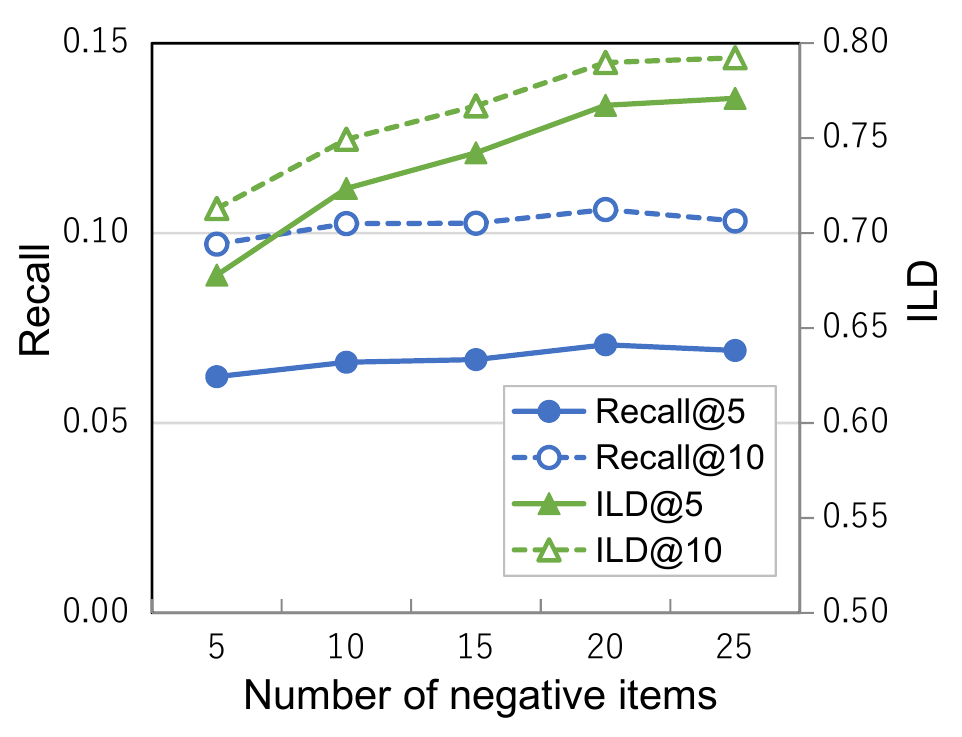}%
	}
	\subfloat[Movielens\label{sfig:movielens_negative}]{%
		\includegraphics[width=0.3\textwidth]{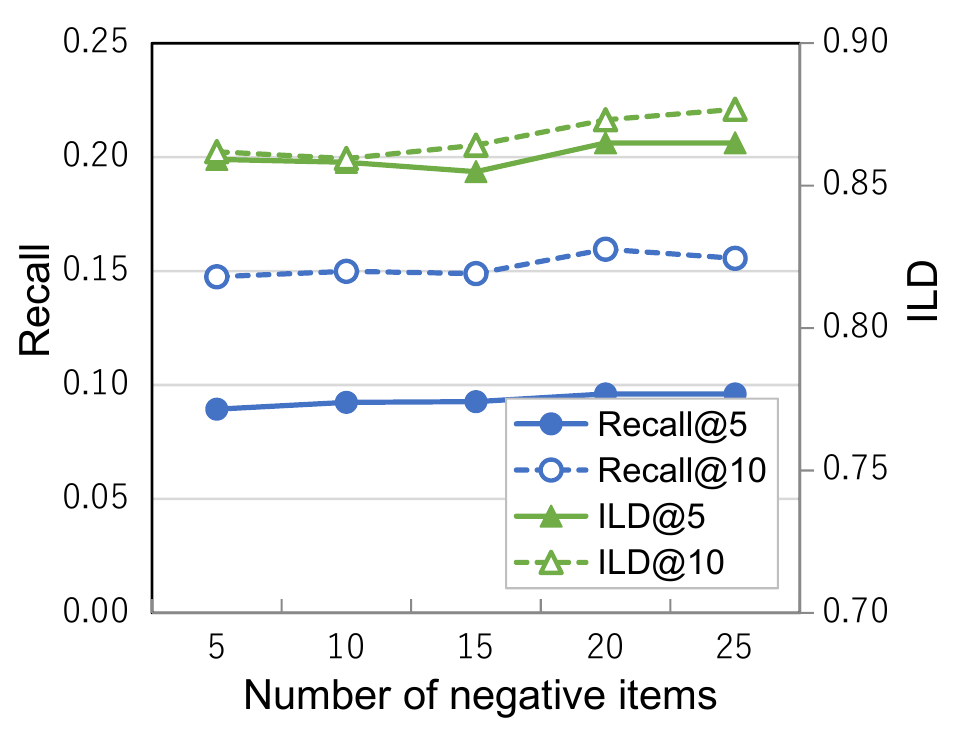}%
	}
	\vspace{-0mm}
	\caption{Impacts  of number of negative samples $P$.}
	\label{fig:neg_variation}
	\vspace{-1mm}
\end{figure*}

\begin{figure*}[htb]
	\vspace{-2mm}
	\centering
	\subfloat[Music\label{sfig:music_aspect}]{%
		\includegraphics[width=0.3\textwidth]{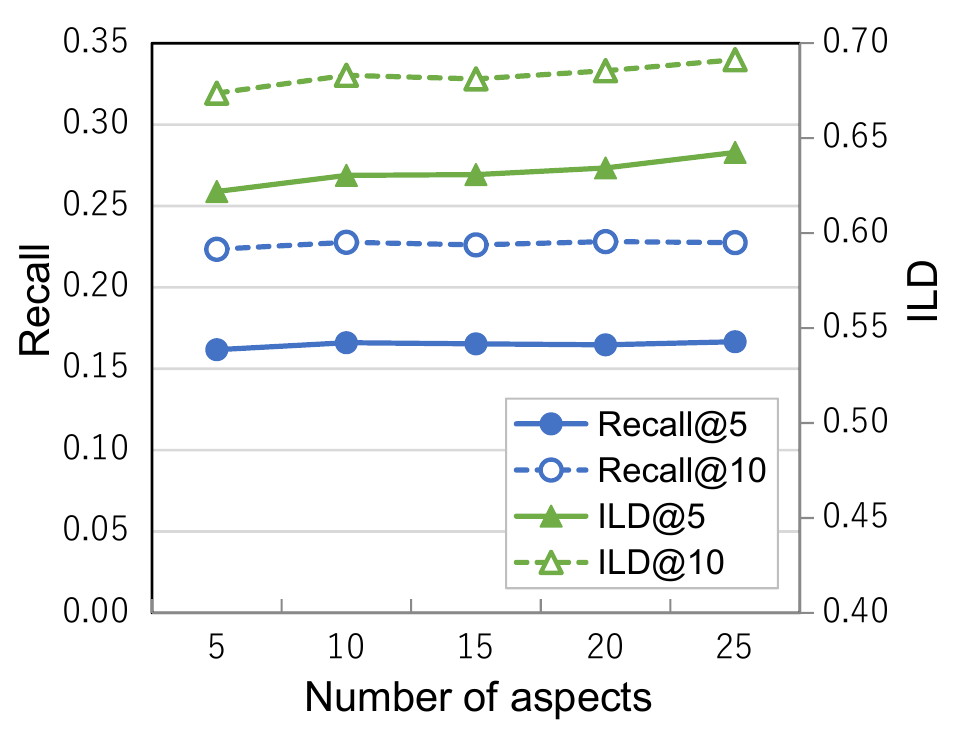}%
	}
	\subfloat[Beauty\label{sfig:beauty_aspect}]{%
		\includegraphics[width=0.3\textwidth]{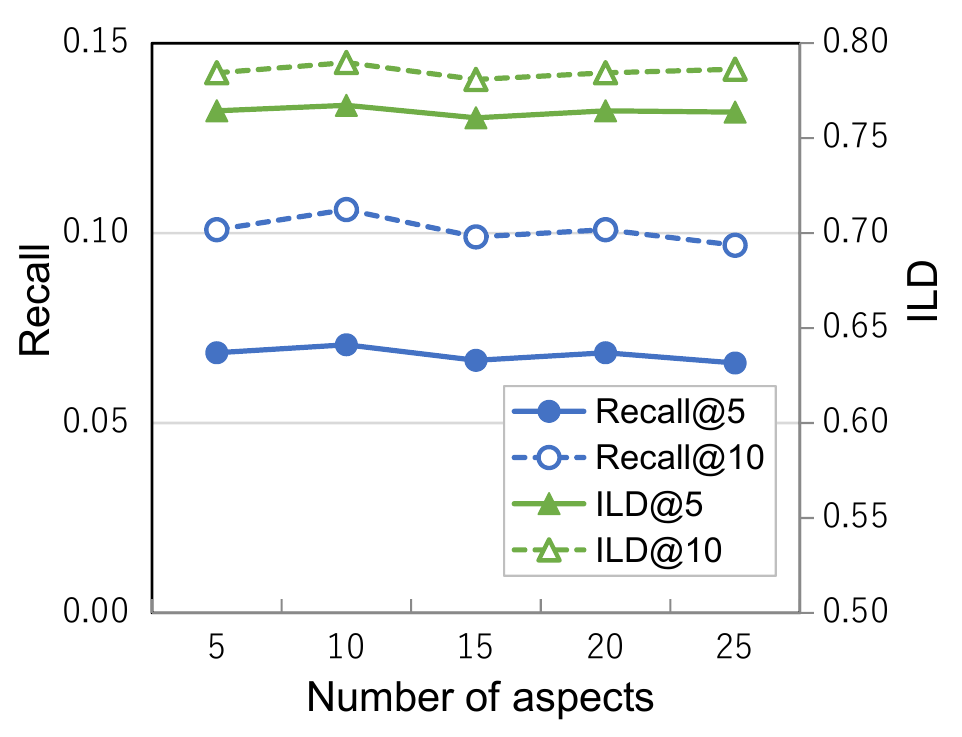}%
	}
	\subfloat[Movielens\label{sfig:movielens_aspect}]{%
		\includegraphics[width=0.3\textwidth]{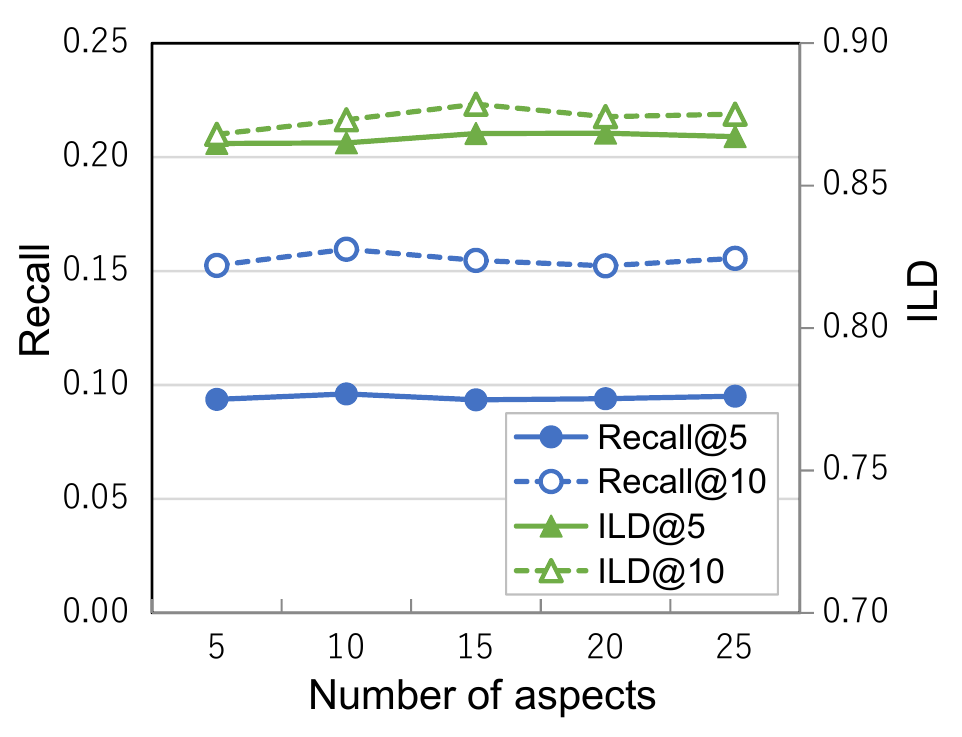}%
	}
	\vspace{-0mm}
	\caption{Impacts  of number of aspects $K$.}
	\label{fig:aspect_variation}
	\vspace{-1mm}
\end{figure*}

\begin{table*}[htb]
	\caption{Ablation results on three datasets}
	\label{tab:ablation_performance}
	\centering
	\small
	\setlength{\tabcolsep}{1.96mm}
	\def\arraystretch{1.05}
	\begin{tabular}{|c|c|c|c|c|c|c|c|c|c|c|c|}
		\hline
		\multirow{2}{*}{Datasets}  & \multirow{2}{*}{Model}      & \multicolumn{4}{c|}{Acurracy}                                         & \multicolumn{4}{c|}{Diversity}                                        & \multicolumn{2}{c|}{Trade-off}    \\ \cline{3-12}
		&                             & Recall@5        & Recall@10       & NDCG@5          & NDCG@10         & ILD@5           & ILD@10          & CC@5            & CC@10           & F1@5            & F1@10           \\ \hline \hline
		\multirow{7}{*}{Music}     & TAML$_{\text{conv}\rightarrow \text{adp}}$ & 0.1636 & 0.2280 & 0.1671 & 0.1934 & 0.6305 & 0.6789 & 0.6014 & 0.7106 & 0.2598 & 0.3414 \\
		& TAML$_{\text{adp}\rightarrow \text{conv}}$ & \textbf{0.1685} & \textbf{0.2327} & \textbf{0.1718} & \textbf{0.1960} & \textbf{0.6412} & \textbf{0.6893} & \textbf{0.6135} & \textbf{0.7168} & \textbf{0.2669} & \textbf{0.3479} \\\cline{2-12}
		& TAML$_\text{conv-only}$              & 0.1554          & 0.2170          & 0.1635          & 0.1862          & 0.6139          & 0.6678          & 0.5903          & 0.6932          & 0.2480          & 0.3276          \\
		& TAML$_\text{adp-only}$ & 0.1383 & 0.1906 & 0.1455 & 0.1659 & 0.5620 & 0.6275 & 0.5364 & 0.6424 & 0.2220 & 0.2924 \\\cline{2-12}
		& TAML$_\text{w/o-dist}$         & 0.1633          & 0.2211          & 0.1707          & 0.1915          & 0.6022          & 0.6518          & 0.6046          & 0.7116          & 0.2569          & 0.3302          \\
		& TAML$_\text{w/o-attn}$            & 0.1586          & 0.2160          & 0.1648          & 0.1901          & 0.6092          & 0.6622          & 0.5962          & 0.7010          & 0.2517          & 0.3257          \\
		& TAML$_\text{w/o-twoway}$ & 0.1560 & 0.2147 & 0.1633 & 0.1865 & 0.6105 & 0.6655 & 0.5927 & 0.7082 & 0.2485 & 0.3247 \\ \hline \hline
		\multirow{7}{*}{Beauty}    & TAML$_{\text{conv}\rightarrow \text{adp}}$ & 0.0659 & 0.1018 & 0.0866 & 0.1019 & 0.7572 & 0.7768 & 0.4876 & 0.6298 & 0.1212 & 0.1800 \\
		& TAML$_{\text{adp}\rightarrow \text{conv}}$ & \textbf{0.0721} & \textbf{0.1071} & \textbf{0.0926} & \textbf{0.1067} & \textbf{0.7675} & \textbf{0.7923} & \textbf{0.5009} & \textbf{0.6445} & \textbf{0.1318} & \textbf{0.1887} \\\cline{2-12}
		& TAML$_\text{conv-only}$            & 0.0672          & 0.0989          & 0.0868          & 0.1004          & 0.7488          & 0.7702          & 0.4833          & 0.6161          & 0.1233          & 0.1753          \\
		& TAML$_\text{adp-only}$    & 0.0591 & 0.0861 & 0.0793 & 0.0907 & 0.6422 & 0.6848 & 0.4290 & 0.5504 & 0.1082 & 0.1530 \\\cline{2-12}
		& TAML$_\text{w/o-dist}$            & 0.0678          & 0.1020          & 0.0865          & 0.1008          & 0.7450          & 0.7705          & 0.4908          & 0.6286          & 0.1243          & 0.1802          \\
		& TAML$_\text{w/o-attn}$           & 0.0652          & 0.1002          & 0.0847          & 0.0995          & 0.7315          & 0.7583          & 0.4951          & 0.6267          & 0.1197          & 0.1770          \\
		& TAML$_\text{w/o-twoway}$ & 0.0670 & 0.0976 & 0.0912 & 0.1037 & 0.7519 & 0.7786 & 0.4917 & 0.6318 & 0.1230 & 0.1735 \\ \hline \hline
		\multirow{7}{*}{Movielens} & TAML$_{\text{conv}\rightarrow \text{adp}}$ & \textbf{0.0975} & \textbf{0.1613} & \textbf{0.6509} & \textbf{0.6684} & 0.8672 & 0.8735 & \textbf{0.3901} & \textbf{0.5651} & \textbf{0.1753} & \textbf{0.2723} \\
		& TAML$_{\text{adp}\rightarrow \text{conv}}$ & 0.0891 & 0.1468 & 0.6262 & 0.6425 & 0.8538 & 0.8606 & 0.3815 & 0.5591 & 0.1614 & 0.2508 \\\cline{2-12}
		& TAML$_\text{conv-only}$              & 0.0945          & 0.1535          & 0.6453          & 0.6600          & 0.8346          & 0.8336          & 0.3834          & 0.5456          & 0.1698          & 0.2593          \\
		& TAML$_\text{adp-only}$  & 0.0684 & 0.1097 & 0.5361 & 0.5574 & \textbf{0.8968} & \textbf{0.9013} & 0.3383 & 0.5299 & 0.1271 & 0.1956 \\\cline{2-12}
		& TAML$_\text{w/o-dist}$            & 0.0951          & 0.1562          & 0.6272          & 0.6472          & 0.8422          & 0.8570          & 0.3779          & 0.5650          & 0.1709          & 0.2642          \\
		& TAML$_\text{w/o-attn}$           & 0.0937          & 0.1531          & 0.6308          & 0.6468          & 0.8517          & 0.8620          & 0.3810          & 0.5551          & 0.1688          & 0.2600          \\
		& TAML$_\text{w/o-twoway}$ & 0.0883 & 0.1447 & 0.6122 & 0.6324 & 0.8577 & 0.8624 & 0.3836 & 0.5618 & 0.1601 & 0.2478 \\ \hline
	\end{tabular}
\end{table*}

In this subsection, we analyze the impacts of hyper-parameters in TAML, including the the dimensionality of the latent vector $D$, the number of negative samples $P$ and the number of aspects $K$. Figure \ref{fig:dim_variation}, Figure \ref{fig:neg_variation}, and Figure \ref{fig:aspect_variation} show the results by varying the dimensionality  in the set of $\{10, 30, 50, 70, 90\}$, tuning the negative sample number amongst $\{5, 10, 15, 20, 25\}$, and varying the number of aspects in the set of $\{5, 10, 15, 20, 25\}$ on three datasets, respectively. Due to the space limitation, we only present the Recall and ILD results.

We first fix $P$ to 20 and $K$ to 20 and vary $D$. We can see from Figure \ref{fig:dim_variation} that, with the increase of the dimensionality of the latent vector, the performance on Recall shows a general upward tendency on all datasets. This indicates that a larger $D$ can capture more latent factors of users and items, which may bring better representation capability. However, if $D$ exceeds a certain value, e.g.70, it will cause overfitting and brings about the decrease of accuracy. We can find a similar trend for diversity on Music and Beauty. The ILD scores on  Movielens ( shown in Figure \ref{fig:dim_variation} (c))  drop first and then rise. This is because most of users on Movielens have wide interests, and thus the model will follow the mainstream when the expression ability is limited or excessive.

We next fix $D$ to 50 and $K$ to 20 and vary $P$. As shown in Figure \ref{fig:neg_variation},  both recall and ILD gradually rise to different extent with the increase of the number of negative samples.  When $P$ is too small, e.g. 5, it is insufficient to achieve optimal performance especially on ILD. The reason is that the limited number of  samples are unable  to describe the border line between the positive and the negative. More negative instances are beneficial for pushing irrelevant and reduplicative items apart to improve diversity. While the number of negative samples reaches at 70, the accuracy performance degrades. It reveals that setting the sampling too large will hurt the performance.

We finally fix $D$ to 50 and $P$ to 20 and vary $K$. We observe from Figure \ref{fig:aspect_variation} that the number of aspects $K$ does not have big impacts on the performance. Indeed, the curves are rather steady with the changing values of $K$. The reason might be that the aspect is a relatively coarse-grained form as category group or user cluster.  In most cases, $K$ = 20 can generate good enough results. Hence we choose this as the setting for K on all datasets. 

\begin{figure*}[!htb]
	\vspace{-3mm}
	\centering
	\subfloat[Music\label{sfig:music_loss}]{%
		\includegraphics[width=0.30\textwidth]{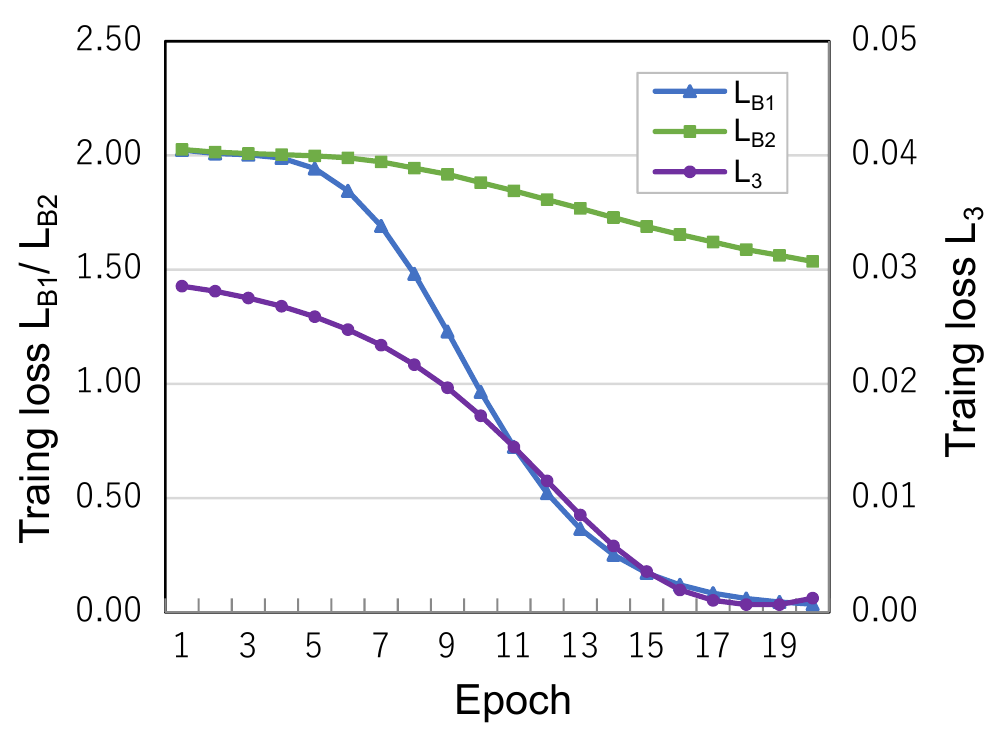}%
	}
	\subfloat[Beauty\label{sfig:beauty_loss}]{%
		\includegraphics[width=0.30\textwidth]{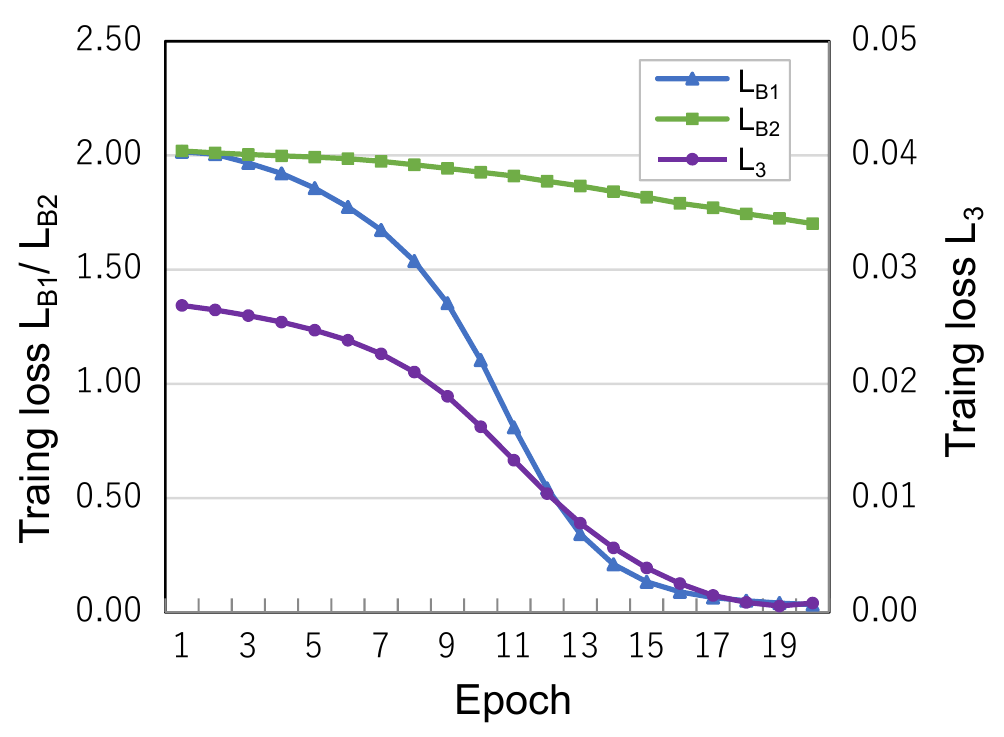}%
	}
	\subfloat[Movielens\label{sfig:movielens_loss}]{%
		\includegraphics[width=0.30\textwidth]{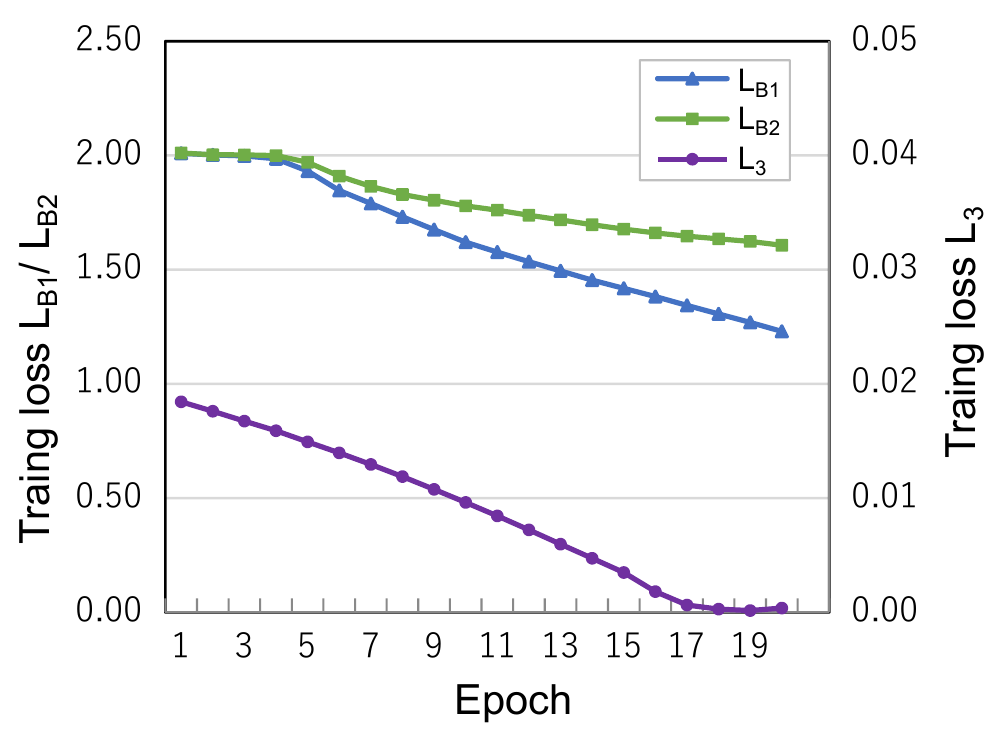}%
	}
	\vspace{-0mm}
	\caption{Training curves of loss $\mathcal{L}_1, \mathcal{L}_2$ and $\mathcal{L}_3$ on different datasets.}
	\label{fig:loss_curve}
	\vspace{-3mm}
\end{figure*}

\subsection{Ablation Studies}
We conduct a series of ablation studies to assess the individual contributions of the components or strategies in our model, including the impacts of the order of learning focus, bilateral branches, and two-way adaptive metric learning. The ablation results on three datasets are shown in Table \ref{tab:ablation_performance}.

\subsubsection{Impacts of the order of learning focus}
As aforementioned in Section \ref{sec:skew}, the order of learning focus decides which branch is carefully trained at the later stage, and different datasets should have different orders to encode the domain level diversity. We determine the order by the skew of the diversity distribution, which is 0.19, 0.14, and 0.92 for Music, Beauty, and Movielens, respectively. Consequently, the training order for Music and Beauty is $\text{adp}\rightarrow \text{conv}$, and $\text{conv}\rightarrow \text{adp}$ for Movielens, where $\text{adp}$ and $\text{conv}$ denotes the adaptive  and conventional learning branch, respectively.

In this experiment, we reverse the order to investigate its impact. Specifically, TAML$_{\text{conv}\rightarrow \text{adp}}$ transfers the learning focus from conventional learning branch to adaptive one (i.e., $\alpha = d_{u} * T / T_{max}$) while TAML$_{\text{adp}\rightarrow \text{conv}}$ is on the contrary (i.e., $\alpha = 1 - d_{u} * T / T_{max}$). It is clear that all the reversed order results in inferior performance. This is consistent with the data biases, i.e., the users in Music and Beauty have focused interests, while a large number of users have broader interests in Movielens. 

\subsubsection{Impacts of bilateral branches}
One key property of our model is the adoption of BBN architecture for balancing accuracy and diversity. To verify this, we remove one branch away for a comparison. TAML$_\text{conv-only}$ and TAML$_\text{adp-only}$ retains conventional learning branch and adaptive learning branch, respectively. The degradation of diversity for TAML$_\text{conv-only}$ reveals that the adaptive branch plays an important role in promoting diversity. Meanwhile, the accuracy performance for TAML$_\text{adp-only}$ drops sharply on all datasets due to the modification of original data distribution. Both these clearly demonstrate the necessity of bilateral branches for two objectives.

\subsubsection{Impacts of two-way adaptive metric learning}
Finally, we investigate the impacts of two-way adaptive metric learning. Specifically, TAML$_\text{w/o-attn}$ and TAML$_\text{w/o-dist}$  remove the attentive matching vector in relevance relation and the diversity relation, respectively from two types of connective relations, and TAML$_\text{w/o-twoway}$ only reserves the commonly used user-to-item direction and removes the item-to-user direction.

The accuracy performance of TAML$_\text{w/o-dist}$ changes slightly but its diversity declines a lot compared to the standard TAML. On the other hand, the downward trend of TAML$_\text{w/o-attn}$ is more obvious on accuracy than diversity. This is consistent with our assumption that attention matching captures the exact relations between the historical behaviors and the target user  while the diversity distribution describes the varying  range of personal interests. Lastly, TAML$_\text{w/o-twoway}$ drops sharply in terms of accuracy and diversity, proving that two-way translation is critical in modeling user-item relationship.

\subsection{Training Process Analysis}

In order to gain more insights into the training process of two branches as well as the tendency of consistent loss, we plot the curves of the model status recorded by training loss of $L_1, L_2$, and $L_3$ in Figure \ref{fig:loss_curve}. As can be seen, the different losses decline quickly first and slow down gradually and finally reach the convergence when the training epoch increases. $L_1$ has a larger decline degree compared with $L_2$, owing to that the samples in $L_1$ are derived from original data distribution while the samples in $L_2$ come from less frequent categories and it is more difficult to optimize $L_2$. Beyond that, the decreasing trend of $L_3$ is similar to that of $L_1$ yet at a small scale. In general, different losses decline steadily, which proves that our model is stable and easy to train.

\subsection{Analysis on Model's Adaptive Ability to Diversity}

\begin{figure*}[!htb]
	\vspace{-3mm}
	\centering
	\subfloat[Music\label{sfig:rec_distribution_music}]{%
		\includegraphics[width=0.80\textwidth]{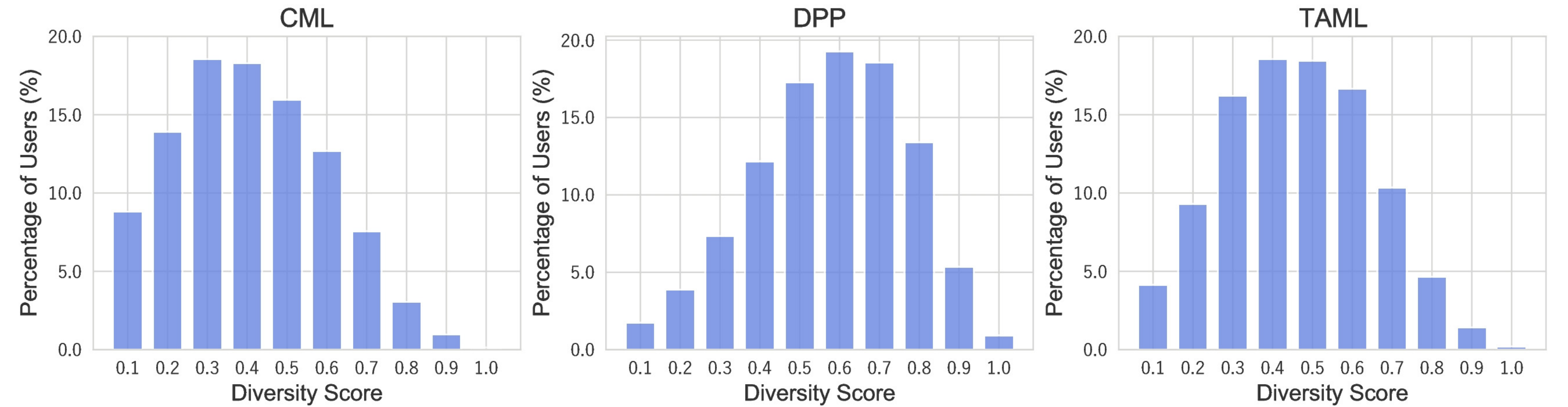}%
	}
	
	\subfloat[Beauty\label{sfig:rec_distribution_beauty}]{%
		\includegraphics[width=0.80\textwidth]{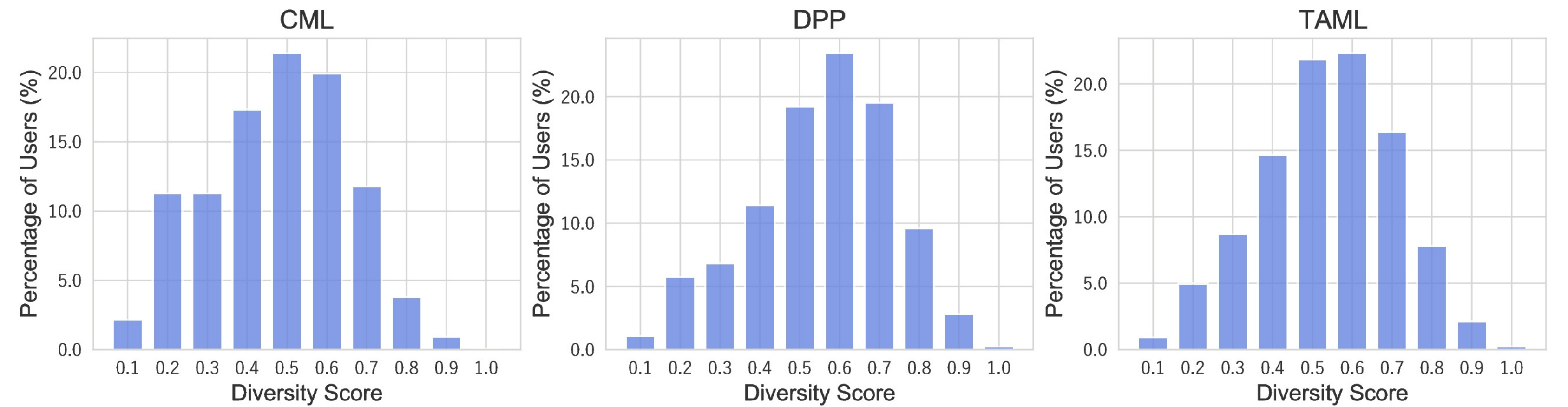}%
	}
	
	\subfloat[Movielens\label{sfig:rec_distribution_movielens}]{%
		\includegraphics[width=0.80\textwidth]{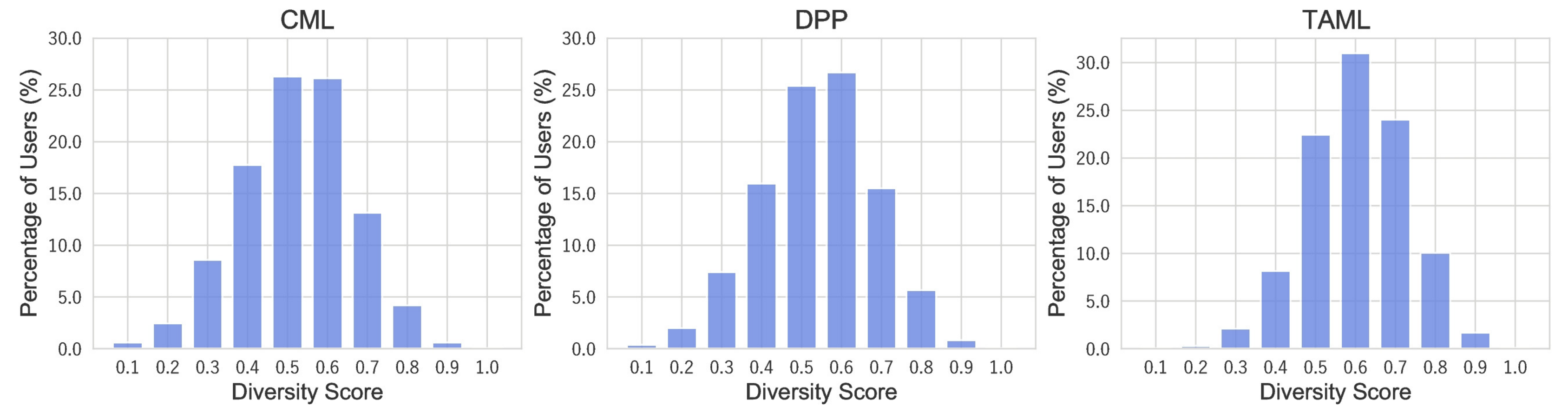}%
	}
	\vspace{-0mm}
	\caption{The distribution of predicted user's diversity on three datasets.}
	\label{fig:rec_dist}
	\vspace{-3mm}
\end{figure*}

\begin{table}[]
	\caption{MSE performance comparison between the predicted user diversity and the original one. }
	\label{tab:rec_dist}
	\centering
	\setlength{\tabcolsep}{3.5mm}	
	\renewcommand\arraystretch{1.2}
	
	\begin{tabular}{|c|c|c|c|}
		\hline
		\diagbox[width=7em]{Datasets}{Models} & CML & DPP & TAML \\ \hline \hline
		Music & 0.0024 & 0.0025 & \textbf{0.0014} \\ \hline
		Beauty & 0.0012 & 0.0013 & \textbf{0.0008} \\ \hline
		Movielens & 0.0111 & 0.0094 & \textbf{0.0059} \\ \hline
	\end{tabular}
\end{table}

In this subsection, we further analyze the adaptive ability of our proposed TAML model on encoding  diversity on different domains. Specifically, we define the predicted diversity score of each user as the number of recommended categories among Top@K items divided by the total number of recommendation K. In order to assess the adaptive ability, we plot the predicted user's diversity distribution conducted by our TAML and that by other two typical baselines, namely CML (accuracy-oriented) and DPP (diversity-promoting) on three datasets.

As shown in Figure \ref{fig:rec_dist},  on different domains, the distribution is presented as left-skewed for CML, and is right-skewed for DPP. The reason is that CML is based on CF and can only cover a small percentage of similar items, which results in its overall low diversity. On the other hand, DPP achieves high diversity using the global trading-off strategy but it ignores the individual difference. In contrast, our TAML model shows different skewed trends on different domains and it yields diversity score distribution close to the original one in Figure \ref{fig0}. It demonstrates that our framework can adapt to different data biases and generate user-specific recommendation lists.

In addition, we examine the difference between the predicted user diversity and the original one. We take the original diversity score defined in Section \ref{sec:userdiver} as ground-truth and calculate the mean square error (MSE) for different methods.  From the results in Table \ref{tab:rec_dist}, it is clear that our TAML achieves the best MSE.  This suggests that our recommended results are mostly consistent with the user's individual diversity in original data and thus can improve the user experience.

\section{Conclusion}
In this paper, we present a novel TAML model for diversified recommendation. It adopts the bilateral branch network as the main architecture for two-objective optimization, i.e., maintaining accuracy in conventional learning branch and increasing diversity in adaptive learning branch. We propose to adaptively balancing the learning procedure between two branches so as to encode the domain-level diversity. We further present a two-way metric learning as the backbone for each branch, which captures the item's orientation towards target users and treats the user level diversity as a special relation connecting the user and the item. Extensive experiments on three real-world datasets demonstrate the effectiveness of our proposed model as well as its distinct components or learning strategies.

\bibliographystyle{ACM-Reference-Format}
\bibliography{b3rec}

\end{document}